\documentclass[journal,twoside,web]{ieeecolor2}
\pdfoutput=1
\usepackage{tmi}
\usepackage{cite}
\usepackage{cuted}
\usepackage{amsmath,amssymb,amsfonts}
\usepackage{algorithm}
\usepackage{algpseudocode}
\usepackage{graphicx}
\usepackage{textcomp}
\usepackage{url}

\usepackage{color}
\definecolor{gray}{rgb}{0.6,0.6,0.6}
\definecolor{red}{rgb}{0.85,0,0}
\definecolor{green}{rgb}{0,0.85,0}
\definecolor{blue}{rgb}{0,0,0.85}
\definecolor{beige}{rgb}{0.92,0.87,0.78}

\def\BibTeX{{\rm B\kern-.05em{\sc i\kern-.025em b}\kern-.08em
    T\kern-.1667em\lower.7ex\hbox{E}\kern-.125emX}}
\DeclareMathOperator{\pos}{pos}
\DeclareMathOperator*{\argmin}{arg\,min}

\def \ylil {y_{\lambda, \, i \ell}}
\def \yblil {\bar{y}_{\lambda, \, i \ell}}
\def \yll {y_{\lambda, \, \ell}}
\def \yl {y_{\lambda}}
\def \ybl {\bar{y}_{\lambda}}
\def \ypl {y^\prime_{\lambda}}
\def \yplil {y^\prime_{\lambda, \, i \ell}}

\def \yml {y_{\mu , \,  \ell}}
\def \ym {y_{\mu}}
\def \ybm {\bar{y}_{\mu}}
\def \ypm {y^\prime_{\mu}}
\def \ypml {y^\prime_{\mu, \, \ell}}
\def \ybml {\bar{y}_{\mu, \, \ell}}

\def \zl {z_{\lambda}}
\def \zpl {z^\prime_{\lambda}}
\def \zppl {z^{\prime\prime}_{\lambda}}
\def \zbl {\bar{z}_{\lambda}}

\def \zm {z_{\mu}}
\def \zpm {z^\prime_{\mu}}

\def \zbm {\bar{z}_{\mu}}

\def \ul {u_{\lambda}}
\def \ulil {u_{\lambda, \, i \ell}}

\def \um {u_{\mu}}

\def \vl {v_{\lambda}}
\def \vm {v_{\mu}}

\def \sl {\sigma_\lambda}
\def \sm {\sigma_\mu}
\def \tl {\tau_\lambda}
\def \tm {\tau_\mu}
\def \gl {\gamma_\lambda}
\def \gm {\gamma_\mu}
\def \kpp {{k^{\prime\prime}}}

\def \nl {\nu_\lambda}
\def \nm {\nu_\mu}

\begin{document}
\title{Simultaneous activity and attenuation estimation in TOF-PET with TV-constrained nonconvex optimization}
\author{Zhimei Ren, Emil Y. Sidky, \IEEEmembership{Member, IEEE}, Rina Foygel Barber,
Chien-Min Kao, \IEEEmembership{Senior Member, IEEE}, and Xiaochuan Pan, \IEEEmembership{Fellow, IEEE}
\thanks{This work is supported in part by
National Science Foundation grants DMS-1654076 and DMS-2023109,
Office of Naval Research grant N00014-20-1-2337, and
NIH grants R01-EB026282, R01-EB023968, R01-EB029948, and R21-CA263660. 
The contents of this article are solely the responsibility of
the authors and do not necessarily represent the official views of the National Institutes of
Health.}
\thanks{Z. Ren is with the Dept. of Statistics and Data Science, University of Pennsylvania (zren@wharton.upenn.edu).}
\thanks{R. F. Barber is with the Dept. of Statistics, University of Chicago (rina@uchicago.edu).}
\thanks{E. Y. Sidky, C.-M. Kao, and X. Pan are with the
Dept. of Radiology, University of Chicago (sidky@uchicago.edu, ckao95@uchicago.edu, xpan@uchicago.edu).}}

\maketitle

\begin{abstract}
An alternating direction method of multipliers (ADMM) framework is developed for nonsmooth biconvex optimization
for inverse problems in imaging. In particular, the simultaneous estimation of activity and attenuation (SAA) problem
in time-of-flight positron emission tomography (TOF-PET) has such a structure when maximum likelihood estimation (MLE) is
employed. The ADMM framework is applied to MLE for SAA in TOF-PET, resulting in the ADMM-SAA algorithm. This algorithm
is extended by imposing total variation (TV) constraints on both the activity and attenuation map, resulting in
the ADMM-TVSAA algorithm. 
The performance of this algorithm is illustrated using the penalized
maximum likelihood activity and attenuation estimation (P-MLAA) algorithm as a reference.
Additional results on step-size tuning and on the use of unconstrained ADMM-SAA are presented in the
previous arXiv submission: arXiv:2303.17042v1.
\end{abstract}


\section{Introduction}
\label{sec:introduction}
\IEEEPARstart{N}{uclear} medicine imaging modalities such as single-photon emission computed tomography (SPECT)
and positron emission tomography (PET) require the input of a gamma ray attenuation map for quantitatively
accurate imaging.  The combination of nuclear medicine imaging with other image modalities such as X-ray
computed tomography (CT) \cite{kinahan1998attenuation,xia2011ultra}
or magnetic resonance imaging (MRI) \cite{burgos2014attenuation} provides a means
for estimating the necessary attenuation map.  There are, however, challenges in the separate
attenuation map estimation. Use of CT-based attenuation maps requires extrapolation of the photon
attenuation map from the diagnostic X-ray energy range to 511 keV and registration of the PET and CT
imaging, which can be particularly difficult in the presence of motion\cite{osman2003respiratory}. The use
of MRI to estimate a synthetic CT image is further complicated by the fact that bone and air have similar
gray values in MRI while bone has a significantly higher attenuation coefficient for gamma rays.

To avoid a separate measurement for obtaining the gamma ray attenuation map, a long-standing inverse problem
of interest has been to simultaneously estimate the attenuation and activity distributions from emission data
alone \cite{natterer1992attenuation,nuyts1999simultaneous}. To address simultaneous activity and attenuation (SAA)
estimation,  Nuyts {\it et al.} \cite{nuyts1999simultaneous} use maximum likelihood
to invert the algebraic SAA model, and they find that accurate activity distributions can be recovered by appropriately
regularizing the attenuation map. The regularization involves the use of Gibbs and intensity priors
on the attenuation distribution that encourage local smoothness and clustering of values around known 
attenuation values for tissues in the scanned subject. Another interesting result for the SAA
problem is obtained in considering time-of-flight positron emission tomography (TOF-PET) \cite{lewellen1998time}.
Defrise {\it et al.} \cite{defrise2012time} exploit an analytic range condition \cite{defrise2008continuous,panin2010restoration}
for the continuous TOF-PET model and obtain a uniqueness
result that the attenuation factor and activity can be determined up to a multiplicative constant.
Returning to the SAA algebraic model for TOF-PET, a comprehensive study of this inverse problem using maximum
likelihood estimation is presented in Rezaei {\it et al.} \cite{rezaei2012simultaneous}, where it is found that
the activity and attenuation maps can be recovered if the timing resolution of the TOF measurements is sufficiently
high and if support constraints are exploited.
We note an intriguing extension of the SAA problem where the background radiation from Lutetium-176, present
in PET scintillators composed of either lutetium oxyorthosilicate (LSO) or lutetium-yttrium orthosilicate (LYSO),
is exploited to provide additional information on the subject's attenuation map without the need for a separate
scan \cite{cheng2020maximum}.
Also, in the context of PET/MRI, anatomical information from standard MRI protocols can be used as a prior to
inform the SAA estimation without the need for dedicated pulse sequences need for MR-based attenuation correction
\cite{salomon2010simultaneous}.

In this work, we seek to build off of Ref. \cite{rezaei2012simultaneous} and develop an image reconstruction framework
for the SAA problem in TOF-PET that can incorporate nonsmooth, convex constraints in the maximum likelihood estimation.
Such constraints can help to achieve stable inversion of the SAA estimation problem.
Of particular interest, here, is the use
of total variation (TV) constraints on both activity and attenuation distributions. We have previously exploited
such constraints in the context of nuclear medicine imaging; in Refs.~\cite{wolf2013few} and \cite{zhang2016investigation}
TV constraints are exploited to enable sparse-data sampling configurations in SPECT and PET, respectively.  In Ref.~\cite{zhang2018optimization},
a similar methodology is used for image reconstruction in low-count list-mode TOF-PET.

The image reconstruction algorithms developed in Refs.~\cite{wolf2013few,zhang2016investigation,zhang2018optimization} are all
instances of a general primal-dual (PD) solver for nonsmooth convex optimization developed by Chambolle and Pock
\cite{chambolle2011first,sidky2012convex}. The optimization problem posed by applying TV-constraints to the SAA estimation problem, however,
is nonsmooth and nonconvex. In our recent work, we develop a framework for such problems in imaging, where the optimization
can be split into convex terms plus differentiable terms that are possibly nonconvex \cite{barber2020convergence}. This
framework is based on the alternating direction method of multipliers (ADMM) \cite{boyd2011distributed} in a way that is
closely related to the PD algorithm. This framework has been successfully
applied to the nonsmooth and nonconvex optimization
problem that arises in spectral computed tomography (CT) when the spectral response of the measurement is included in
the data model \cite{schmidt2022addressing}. Here, we modify this framework to address biconvex optimization
and apply it to the SAA estimation problem
with convex constraints. The SAA data model and imaging problem are specified in Sec. \ref{sec:theory}, where we then
develop an ADMM algorithm to solve the associated optimization problem. As the focus of this work is mainly on the SAA
inverse problem, we conduct a number of studies on noiseless TOF-PET data in Sec. \ref{sec:results} that explore
the range of TOF-PET parameters that allow exact recovery of activity and attenuation factors. Also presented in this
section are results with noisy data that demonstrate the stability of the proposed algorithm. In Sec.~\ref{sec:conclusion}
the results are discussed and the conclusions of the work are given.

\section{Image reconstruction model and algorithms}
\label{sec:theory}

In presenting the SAA algorithm TOF-PET, we consider a two dimensional (2D) simulation where the lines-of-response (LORs)
are organized in parallel-ray fashion and are specified in the same way that the 2D Radon transform is parameterized.
For the TOF-PET model, the Radon transform is modified by including weighted line-integration that accounts for TOF
information that helps to localize the positron-electron annihilation along a given LOR. After specifying the TOF-PET
data model, the MLAA algorithm from Rezaei {\it et al.} \cite{rezaei2012simultaneous}
is briefly summarized. We then present the nonconvex ADMM
algorithm that performs SAA estimation with non-smooth convex constraints.

\subsection{TOF-PET modelling}

The measurement model for the mean data in TOF-PET is
\begin{equation}
\label{TOFPETmodel}
c_{i \ell} = \exp \left[ - P^\top_{\ell} \mu \right] \cdot T^\top_{i \ell} \lambda,
\end{equation}
where $\lambda$ and $\mu$ are the unknown activity and attenuation maps, respectively;
$T_{i \ell }$ is the TOF sensitivity
image for TOF window $i$, LOR $\ell$; $P_{\ell }$ is the X-ray projection matrix sensitivity image for LOR $\ell$.
For defining the TOF projection matrix $T$, the TOF window sensitivity along the LOR is specified as
\begin{equation*}
w_i(t) = \exp [- (t - t_i)^2 / (2 \sigma_\text{TOF}) ],
\end{equation*}
where the sampling along the LOR is half of the full-width-half-maximum (FWHM) of this Gaussian distribution
\begin{equation*}
\Delta t =t_{i+1} - t_i = \text{FWHM}/2 = \sqrt{2 \log 2} \cdot \sigma_\text{TOF}.
\end{equation*}
For this work, scatter coincidences and random events are not considered.

\subsection{Imaging model based on nonconvex optimization}

We consider performing SAA using likelihood maximization, where the measured coincidence count data are assumed to
follow a multivariate, mutually independent Poisson distribution
\begin{equation*}
C_{i \ell} \sim \text{Poisson}(c_{i \ell}).
\end{equation*}
Equivalently, this estimation is performed by minimization of the negative log-likelihood,
\begin{align}
&l(\lambda,\mu)= \sum_{i \ell} \left\{ c_{i \ell} - C_{i \ell} \cdot \log c_{i \ell} \right\} = \label{nll} \\
&\sum_{i \ell} \left\{ \exp(-P^\top_\ell \mu) \cdot T^\top_{i \ell} \lambda  -
C_{i \ell} \cdot (-P^\top_{\ell} \mu + \log(T^\top_{i \ell} \lambda)) \right\}.
\notag
\end{align}

The optimization problem of interest is
\begin{equation}
\label{opt1}
\lambda,\mu = \argmin_{\lambda,\mu} \left\{ l(\lambda,\mu) \; \; | \; \; \mathbf{1}^\top \lambda = N_\text{total}
\text{, } \lambda,\mu \ge 0 \right \}, 
\end{equation}
where $l$ is the negative log-likelihood in Eq.~(\ref{nll}); $\mathbf{1}$ is a vector of size $\lambda$ with unit entries
so that $\mathbf{1}^\top \lambda$ is equivalent to summation over $\lambda$; and $N_\text{total}$ is the total 
number of annihilations.
The constraint on the total number of annihilations is used to overcome the constant ambiguity in the SAA estimation problem
\cite{defrise2012time}. This constraint is enforced in this work instead of the object support constraint investigated in
Rezaei {\it et al.} \cite{rezaei2012simultaneous}.

\subsection{Summary of MLAA}
\label{sec:mlaa}
To solve this imaging model, Rezaei {\it et al.} \cite{rezaei2012simultaneous} developed the MLAA
algorithm. For completeness, we write the MLAA update steps including a minor modification in
Eq.~(\ref{mlaa3}) that 
accommodates the constraint on the total number of annihilations:
\begin{align}
a_{\ell} &= \exp \left[ - \sum_k P_{\ell k} \mu_k \right]\;\;\; \forall \ell, \label{mlaa1}\\
\lambda_k &\leftarrow \frac{\lambda_k}{\sum_{i \ell} a_{\ell} T_{i \ell k}}
\sum_{i \ell} \left\{  T_{i \ell k}
\left( \frac{C_{i \ell}}{\sum_{k^\prime}T_{i \ell k^\prime}\lambda_{k^\prime}} \right)  \right\} \;\;\; \forall k, \label{mlaa2} \\
\lambda &\leftarrow \lambda \left( \frac{N_\text{total}}{\sum_k \lambda_k} \right), \label{mlaa3} \\
\mu_k &\leftarrow \mu_k +\frac{\sum_{i \ell k^\prime} P_{\ell k}  \left( a_\ell T_{i \ell k^\prime} \lambda_{k^\prime}  -  C_{i \ell} \right)}
                          {\sum_{i \ell k^\prime}  P_{\ell k^\prime} P_{\ell k}  a_\ell \sum_{\kpp} T_{i \ell \kpp} \lambda_\kpp} \;\;\; \forall k ,
\label{mlaa4}\\
\mu_k &\leftarrow \pos (\mu_k) \; \; \; \forall k. \label{mlaa5} 
\end{align}
The MLAA algorithm essentially alternates between updating $\lambda$ with a Poisson likelihood EM step and $\mu$
with a Poisson transmission likelihood optimization step. In this MLAA implementation the extra update step in Eq.~(\ref{mlaa3})
enforces the constraint on the total number of annihilations,
and Eq.~(\ref{mlaa5}) performs non-negativity projection, where negative values of $\mu$ are set to zero.
For MLAA the activity $\lambda$ should have a strictly positive initialization, and
this quantity will remain non-negative during the iteration.

Early stopping of the iteration is the primary means of performing regularization with MLAA, but explicit regularization can also
be included with the use of Gibbs smoothing \cite{lange1990convergence,heusser2017mlaa,mehranian2017mr}. In this work, we develop a framework
for SAA which can include nonsmooth regularization.

\subsection{ADMM for nonsmooth and biconvex optimization}

The general convex optimization problem that ADMM solves takes the form
\begin{equation*}
\min_{x,y} \left\{f(x)+g(y)\; \; | \; \; Ax + By =c \right\},
\end{equation*}
where $f$ and $g$ are convex and possibly non-smooth functions; $A$ and $B$ are linear operators;
$x$, $y$ and $c$ are vectors.
The steps of the ADMM algorithm are
\begin{align}
x &\leftarrow \argmin_{x^\prime} \Bigl\{ f(x^\prime)  +u^\top A x^\prime  \Bigr. \notag \\
& \Bigl.  +\tfrac{1}{2}\|Ax^\prime+By-c\|^2_\Sigma +\tfrac{1}{2}\|x^\prime-x\|^2_{H_f} \Bigr\} \label{admm1} \\
y &\leftarrow \argmin_{y^\prime} \Bigl\{ g(y^\prime)  +u^\top B y^\prime \Bigr. \notag \\
&+ \Bigl. \tfrac{1}{2}\|Ax+By^\prime-c\|^2_\Sigma +\tfrac{1}{2}\|y^\prime -y\|^2_{H_g}\Bigr\} \label{admm2}\\
u &\leftarrow  u + \Sigma (A x + B y -c), \label{admm3}
\end{align}
where $\Sigma$, $H_f$, and $H_g$ are symmetric positive definite, and
$\|v\|^2_M \equiv v^\top M v$ for any symmetric positive definite matrix $M$.
Because optimizing the TOF-PET likelihood for SAA is a non-convex optimization problem, the ADMM algorithm
does not directly apply. 
One strategy to adapt ADMM to SAA is to base the ADMM steps on a series of successive convex
approximations as developed by Chun {\it et al.} \cite{chun2016joint}
using the separable quadratic surrogates (SQS) method.
In the present work, we develop an alternative form of ADMM that directly applies to SAA,
exploiting the biconvex structure of the TOF-PET likelihood function;
i.e. fixing
either $\lambda$ or $\mu$, the likelihood is a convex function in the other variable.

The ADMM algorithm can be modified to accommodate a biconvex function, and we consider the
case that only $g$ is a biconvex function
\begin{equation*}
g(y) = g(y_1,y_2),
\end{equation*}
where $y$ is the concatenation of $y_1$ and $y_2$;
and $g(y_1,\cdot)$ and $g(\cdot,y_2)$ are convex functions for fixed $y_1$ and $y_2$, respectively.
To accommodate the biconvexity of $g$, the second update equation, Eq.~(\ref{admm2}), is replaced by
an inner iteration with the following update equations
\begin{align}
y_1 &\leftarrow \argmin_{y^\prime_1} \Bigl\{ g(y^\prime_1,y_2) + u^\top B \, (y^\prime_1,y_2) \Bigr. \label{iadmm1}\\
+& \Bigl. \tfrac{1}{2}\|Ax + B \, (y^\prime_1,y_2) - c\|^2_\Sigma +\tfrac{1}{2}\|(y^\prime_1,y_2) - (y_1,y_2)\|^2_{H_g}\Bigr\} \notag \\
y_2 &\leftarrow \argmin_{y^\prime_2} \Bigl\{ g(y_1,y^\prime_2)  + u^\top B \, (y_1,y^\prime_2) \Bigr. \label{iadmm2}\\
 +& \Bigl. \tfrac{1}{2}\|Ax + B \, (y_1,y^\prime_2)-c\|^2_\Sigma +\tfrac{1}{2}\|(y_1,y^\prime_2)-(y_1,y_2)\|^2_{H_g}\Bigr\}. \notag
\end{align}
The inner loop consists of alternating between Eqs.~(\ref{iadmm1}) and (\ref{iadmm2}) for a predetermined number of iterations $N_y$,
where $N_y \ge 1$.
After the inner loop is completed, the ADMM iteration continues
with Eq.~(\ref{admm3}) after the following assignment
\begin{equation*}
y = (y_1, \; y_2).
\end{equation*}
This inner loop, specified in Eqs.~(\ref{iadmm1}) and (\ref{iadmm2}), is computationally efficient if multiplication
by the matrix $B$ is efficient; this is the case in our application because we consider $B=I$ where $I$ is the identity matrix.
Note that multiplication by $A$ is not performed within this inner iteration because the matrix $A$ only appears
in the term $Ax$ which is computed before entering the inner loop.

\subsection{ADMM for large-scale tomographic image reconstruction}
For the large-scale optimization problems that arise in tomographic image reconstruction, the update step
in Eq.~(\ref{admm1}) can be problematic because of the term $Ax$, which appears in the minimization over $x$.
The matrix $A$ usually contains the system matrix for the imaging model, and computation of $Ax$ can be expensive
particularly for 3D imaging; thus numerical solution of Eq.~(\ref{admm1}) may not be feasible. This
``expensive inner loop'' problem can be circumvented by linearization, i.e. by including the additional term
$\tfrac{1}{2} \|x^\prime - x \|^2_{H_f}$ in Eq.~(\ref{admm1}) \cite{barber2020convergence,nien2014fast},
resulting in an algorithm closely related to the primal-dual (PD) algorithm of Chambolle and Pock \cite{chambolle2011first,sidky2012convex}.
Considering only scalar step size parameters, i.e.
\begin{equation*}
\Sigma = \sigma I,
\end{equation*}
the metric $H_f$ in Eq.~(\ref{admm1}) is set to
\begin{equation}
\label{hmetric}
H_f = I/\tau - \sigma A^\top A.
\end{equation}
This choice cancels the $Ax^\prime$ term in Eq.~(\ref{admm1}), and the requirement that $H_f$ be positive definite
yields a constraint on the step sizes $\sigma$ and $\tau$.
In the context of the image reconstruction problem, we also have
\begin{equation*}
H_g=0; \; \; B=-I; \; \; c=0.
\end{equation*}

The ADMM generic optimization problem becomes
\begin{equation}
\label{ADMMopt}
\min_{x,y} \left\{f(x)+g(y) \; | \; Ax - y =0 \right\},
\end{equation}
and the algorithm for convex optimization is then specified by the following
update equations
\begin{align}
x &\leftarrow \argmin_{x^\prime} \Bigl\{ f(x^\prime) + {x^\prime}^\top A^\top(u + \sigma(Ax- y))   \Bigr. \notag \\
& \Bigl.   + \tfrac{1}{2 \tau}\|x^\prime-x\|^2 \Bigr\} \label{nadmm1} \\
y &\leftarrow \argmin_{y^\prime} \Bigl\{ g(y^\prime)  -u^\top y^\prime
+ \tfrac{\sigma}{2}\|Ax- y^\prime \|^2 \Bigr\} \label{nadmm2}\\
u &\leftarrow  u + \sigma (A x -  y). \label{nadmm3}
\end{align}
Aside from minor details, this set of update equations is equivalent to the PD algorithm,
but as a starting point to modify the update steps for non-convex optimization, this form
is more convenient because both $f$ and $g$ functions appear directly in the updates.
In contrast, the PD algorithm dualizes $g$ and the convex conjugate $g^\star$ is needed.
If it is desired to apply PD to non-convex $g$, figuring out what to put in place
of $g^\star$, while possible\cite{barber2016},
adds another layer of complication to the algorithm development.

The modification of the linearized ADMM updates for addressing the case where $g$ is biconvex
replaces Eq.~(\ref{nadmm2}) with inner loop update equations
\begin{align}
y_1 &= \argmin_{y^\prime_1} \Bigl\{ g(y^\prime_1,y_2)  -u^\top (y^\prime_1,y_2)  \Bigr. \label{inadmm1}\\
&+ \Bigl. \tfrac{\sigma}{2}\|Ax - (y^\prime_1,y_2) \|^2 \Bigr\} \notag \\
y_2 &= \argmin_{y^\prime_2} \Bigl\{ g(y_1,y^\prime_2)  - u^\top (y_1,y^\prime_2) \Bigr. \label{inadmm2}\\
& + \Bigl. \tfrac{\sigma}{2}\|Ax - (y_1,y^\prime_2)\|^2 \Bigr\}. \notag
\end{align}
Convergence of this modified ADMM algorithm for biconvex functions is not theoretically guaranteed and thus
convergence is demonstrated empirically.

\subsection{ADMM for SAA in TOF-PET}
\label{sec:ADMM-SAA}

The instantiation of ADMM for SAA estimation by minimization of the negative log-likelihood
is covered here in detail. The optimization problem of interest,
restated from Eq.~(\ref{opt1}), is
\begin{equation}
\label{opt2}
\lambda,\mu = \argmin_{\lambda,\mu} \left\{ l(\lambda,\mu) \:\; | \:\; \mathbf{1}^\top \lambda = N_\text{total}, \; \; \lambda \ge 0,
\; \; \mu \ge 0 \right\}.
\end{equation}
In this sub-section, we map this optimization problem on to the ADMM algorithm, derive the
$x$-update and biconvex $y$-updates,
and provide the pseudo-code for SAA estimation.

To map the optimization problem in Eq.~(\ref{opt2}) onto the generic ADMM optimization in Eq.~(\ref{ADMMopt}),
the primal, splitting, and dual variables $x$, $y$, and $u$, are respectively assigned as
\begin{equation*}
x = \left(
\begin{array}{c}
\lambda \\
\mu
\end{array}
\right),  \; \;
y = \left(
\begin{array}{c}
y_\lambda \\
y_\mu
\end{array}
\right), \; \;
u = \left(
\begin{array}{c}
u_\lambda \\
u_\mu
\end{array}
\right).
\end{equation*}
The linear system $A$ is assigned as
\begin{equation*}
A = \left(
\begin{array}{cc}
T & 0\\
0 & P
\end{array}
\right).
\end{equation*}
The convex function $f$ is used to represent the
non-negativity constraints and the constraint on the total number of annihilations
by setting
\begin{equation}
\label{fdef}
f(\lambda,\mu) = \delta( \mathbf{1}^\top \lambda = N_\text{total}) + \delta (\lambda \ge 0) + \delta (\mu \ge 0),
\end{equation}
where $\delta$ is the convex indicator function, which is zero if the conditional argument is true and
infinity otherwise.
The biconvex function $g$ accounts for the negative log-likelihood objective function in Eq.~(\ref{opt2})
\begin{align}
\label{gdef}
g(\yl,\ym) &= L(\yl,\ym), \\
L(\yl,\ym) &= \sum_{i \ell} \Bigl\{ \exp(-\yml) \cdot \ylil \Bigr. \notag\\
& \Bigl. - C_{i \ell}  \cdot (-\yml + \log(\ylil)) \Bigr\},
\notag
\end{align}
where 
\begin{equation*}
l(\lambda,\mu) =  L(T \lambda, P \mu).
\end{equation*}

\subsubsection*{Parametrization of the step sizes}
Step size selection is a critical issue for first-order, large-scale optimization
algorithms. There can be much flexibility in the step size selection, and it is important
to select a minimal set of free parameters that are effective for algorithm efficiency
but not too cumbersome in the tuning procedure.
Because the system matrix $A$ for SAA is block-diagonal, a slight generalization of
the ADMM linearization is considered. The metric $H_f$ is written as
\begin{align*}
H_f =& \left(
\begin{array}{cc}
H_\lambda & 0\\
0 & H_\mu
\end{array}
\right), \\
H_\lambda =& \frac{I}{\tau_\lambda} - \sigma_\lambda T^\top T, \\
H_\mu =& \frac{I}{\tau_\mu} - \sigma_\mu P^\top P,
\end{align*}
and the step size parameters are chosen according to
\begin{equation*}
\sigma_\lambda \tau_\lambda = 1/\| T \|^2_2, \; \;
\sigma_\mu \tau_\mu = 1/\| P \|^2_2,
\end{equation*}
where $\| M \|_2$ is the largest singular value of the matrix $M$.
With four step size parameters and two equality constraints, there are two free step size
parameters. Specifically, the step size ratios, $\rho_\lambda$
and $\rho_\mu$, are chosen
to be the free parameters that need to be tuned:
\begin{align}
\sigma_\lambda =& \rho_\lambda/\|T\|_2, \; \tau_\lambda = 1/(\rho_\lambda\|T\|_2), \label{stepl}\\
\sigma_\mu=& \rho_\mu/\|P\|_2, \; \tau_\mu = 1/(\rho_\mu\|P\|_2). \label{stepm}
\end{align}
Tuning of $\rho_\lambda$ and $\rho_\mu$ is a necessary step any time the $T$ or $P$
matrices are changed due to, for example, a change in scan configuration or sampling pattern.

\subsubsection*{The $x$-update}
For the SAA problem in TOF-PET the $x$-update in Eq.~(\ref{nadmm1}) splits into two optimization problems
\begin{align}
\lambda &\leftarrow \argmin_{\lambda^\prime}
\Bigl\{{\lambda^\prime}^\top T^\top(u_\lambda + \sigma_\lambda(T \lambda- y_\lambda))   \Bigr. \label{lamopt} \\
& \Bigl.   + \tfrac{1}{2 \tau_\lambda}\|\lambda^\prime-\lambda\|^2 \; \; | \; \;
\mathbf{1}^\top \lambda^\prime = N_\text{total}\text{,  } \lambda^\prime \ge 0 \Bigr\}, \notag \\
\mu&\leftarrow \argmin_{\mu^\prime}
\Bigl\{{\mu^\prime}^\top P^\top(u_\mu+ \sigma_\mu(P \mu- y_\mu))   \Bigr.  \label{muopt} \\
& \Bigl.   + \tfrac{1}{2 \tau_\mu}\|\mu^\prime-\mu\|^2 \; \; | \; \; \mu^\prime \ge 0 \Bigr\}, \notag
\end{align}
where the indicator terms of the convex function $f$ from Eq.~(\ref{fdef}) are incorporated 
as constraints in the $\lambda$- and $\mu$-update equations.
The optimization problem for the $\mu$-update in Eq.~(\ref{muopt})
is solved by setting the gradient of the objective function
to zero and solving for $\mu^\prime$,
followed by a non-negativity projection to enforce the constraint on $\mu_\prime$
\begin{align}
\mu & \leftarrow \mu - \tau_\mu  P^\top \bar{u}_\mu,  \label{mupdate} \\
\bar{u}_\mu & = u_\mu+ \sigma_\mu(P \mu- y_\mu),  \notag \\
\mu & \leftarrow \pos ( \mu), \notag
\end{align}
where the function $\pos (\cdot)$
thresholds negative components of the argument to zero.

For the $\lambda$ optimization problem in Eq.~(\ref{lamopt}),
completing the square in the quadratic objective function,
rescaling the objective function, and ignoring
the $\lambda^\prime$-independent term yields
\begin{align}
\lambda &\leftarrow \argmin_{\lambda^\prime}
\Bigl\{ \tfrac{1}{2}
\left\|\lambda^\prime -
\left( \lambda - \tau_\lambda T^\top \bar{u}_\lambda \right) \right\|^2 \Bigr. \label{lamopt2} \\ 
& \Bigl.  \; \; | \; \;
\mathbf{1}^\top \lambda^\prime = N_\text{total} \text{,  } \lambda^\prime \ge 0 
\Bigr\}, \notag \\
\bar{u}_\lambda & = u_\lambda + \sigma_\lambda(T \lambda- y_\lambda). \notag
\end{align}
This optimization problem is now in the form of a projection onto the positive
simplex, for which an efficient algorithm is developed by Duchi {\it et al.} \cite{duchi2008efficient}.
Computationally, the updates in $\lambda$ and $\mu$ are the most expensive steps in the ADMM algorithm
because they involve forward- and back-projection of $\mu$ and $\bar{u}_\mu$, respectively,
in addition to TOF forward- and back-projection of $\lambda$ and $\bar{u}_\lambda$, respectively.

\subsubsection*{The biconvex $y$-updates}
The $g$ function in Eq.~(\ref{gdef}) is biconvex in that it is convex in $y_\lambda$ if $y_\mu$
is fixed and {\it vice versa}. Splitting up the $g$ function over the two update equations
in Eqs.~(\ref{inadmm1}) and (\ref{inadmm2}) yields
\begin{multline}
\yl = \argmin_{\ypl} \Bigl\{ \sum_{i \ell} \left(
\exp (-\yml) \cdot \yplil  - C_{i \ell} \cdot \log ( \yplil ) \right)  \Bigr. \\
\Bigl.  - \ul^\top \ypl
+ \frac{\sl}{2 } \| \ypl - T\lambda \|^2\; \; | \; \; \ypl \ge 0
\Bigr\}, \label{yoptlam}
\end{multline}
and
\begin{multline}
\ym = \argmin_{\ypm} \Bigl\{ \sum_{i \ell} \left(
\exp (-\ypml) \cdot \ylil  + C_{i \ell} \cdot \ypml   \right)  \Bigr. \\
\Bigl. - \um^\top \ypm
+ \frac{\sm}{2 } \| \ypm - P\mu\|^2 \; \; | \; \; \ypm \ge 0
\Bigr\}, \label{yoptmu}
\end{multline}
noting that the $\exp(-\ym) \cdot \yl$ term is the only one that mixes
the $\yl$ and $\ym$ variables and is therefore common to both
minimization problems.
In order for the biconvex alternation to converge it is necessary to introduce the non-negativity
constraints on $\ypl$ and $\ypm$. Physically, these constraints are redundant with the non-negativity
constraints imposed on $\lambda$ and $\mu$; if these physical constraints are not used, it is still necessary
to impose non-negativity constraints on $T\lambda$ and $P\mu$.

The minimization problems for the $y$-update are both separable over the components
of $\ypl$ and $\ypm$. The minimization over $\ypl$ in Eq.~(\ref{yoptlam})
is solved analytically by setting
the gradient of the objective function to zero,
yielding a quadratic equation when $C_{i \ell} >0$,
\begin{align*}
& \sl \ylil^2  -  b_{i \ell} \ylil  -  C_{i \ell} =0, \\
\text{where  } &  b_{i \ell} = \ulil + \sl T^\top_{i \ell} \lambda - \exp(-\yml),
\end{align*}
and a linear equation when $C_{i \ell} = 0$,
\begin{equation*}
\sl \ylil  -  b_{i \ell}  =0.
\end{equation*}
For the $C_{i \ell} > 0$ case, the non-negativity constraint on $\ypl$ is respected
by selecting the non-negative root of the corresponding quadratic equation, and for
the $C_{i \ell} = 0$ case, the non-negativity constraint on $\ypl$ yields an update,
\begin{equation*}
\ylil = \max ( b_{i \ell}/\sl , 0 ).
\end{equation*}
Both the linear and quadratic cases can be merged
into the following update equation for $\ylil$,
\begin{equation}
\ylil = \left( b_{i \ell} + \sqrt{b^2_{i \ell} + 4 \sl C_{i \ell}} \right)/(2\sl). \label{ylupdate}\\
\end{equation}

Solving the minimization over $\ypm$ in Eq.~(\ref{yoptmu}) is more involved
because setting the gradient of the objective function to zero results in
a transcendental equation, which requires the use of a numerical solver.
The objective function is convex in $\ypm$ and its derivatives are easily
computed analytically. Thus Newton's algorithm can be applied to obtain
an efficient and accurate solution to Eq.~(\ref{yoptmu}). Both the first and
second derivatives of the objective function are needed for Newton's algorithm.
Defining $\psi$ to be the objective function of Eq.~(\ref{yoptmu})
\begin{multline*}
\psi(\ypm) = \sum_{i \ell} \left(
\exp (-\ypml) \cdot \ylil  + C_{i \ell} \cdot \ypml   \right) \\
- \um^\top \ypm + \frac{\sm}{2 } \| \ypm - P\mu\|^2 ,
\end{multline*}
the first derivative of $\psi$ is
\begin{multline}
\frac{\partial \psi(\ypm)}{\partial \ypml} = -\exp (-\ypml) \cdot \yll \label{psi1} \\
+ C_{\ell} - \um + \sm (\ypml - P^\top_\ell \mu), 
\end{multline}
where
\begin{equation*}
\yll = \sum_i \ylil \, , \; \; C_\ell = \sum_i C_{i \ell} \, .
\end{equation*}
The second derivative of $\psi$ is
\begin{equation}
\frac{\partial^2 \psi(\ypm)}{\partial {\ypml}^2} = \exp (-\ypml) \cdot \yll
+ \sm \, , \label{psi2}
\end{equation}
which is strictly positive. Thus Newton's algorithm can be applied without any difficulties
with the following update equation
\begin{equation}
\label{newt}
\ypml \leftarrow \ypml - \frac{\partial \psi(\ypm)}{\partial \ypml}
\left( \frac{\partial^2 \psi(\ypm)}{\partial {\ypml}^2}\right)^{-1}.
\end{equation}
There is also the non-negativity constraint in Eq.~(\ref{yoptmu}), and this
can be accounted for by thresholding negative values of $\ypml$
to zero after the Newton iteration is completed. 

The proposed $y$-update involves two additional levels of iteration.
The first additional level of iteration involves alternating between solving
Eqs.~(\ref{yoptlam}) and (\ref{yoptmu}). In the second additional level 
of iteration Eq.~(\ref{yoptmu}) is solved with the Newton iteration in
Eq.~(\ref{newt}). Nevertheless, these additional nested iterations
do not negatively impact the efficiency of the overall algorithm because
all of the iterations for the $y$-update separate over the components of $y$.
The complete $y$-update computation takes less effort than computing $T\lambda$, the TOF data
of an estimate of the activity map, $\lambda$.
This is one of the useful aspects of the powerful splitting technique that
ADMM exploits.

\begin{algorithm}
\hrulefill
\begin{algorithmic}[1]
\For {$k \leftarrow 1$, $N_\text{iter}$}
   \State $\bar{\lambda}  = T^\top (u_\lambda + \sigma_\lambda (\ybl - y_\lambda))$  \label{p:tback}
   \State $\lambda \leftarrow \mathcal{P}_{\text{simplex}(N_\text{total})}
                               \left(\lambda - \tau_\lambda \bar{\lambda} \right)$ \label{p:lupdate}
   \State $\ybl = T\lambda$ \label{p:t} 
   \State $\bar{\mu}  = P^\top( u_\mu+ \sigma_\mu(\ybm - y_\mu))$ \label{p:pback}
   \State $\mu  \leftarrow \pos (\mu - \tau_\mu  \bar{\mu}) $ \label{p:mupdate}
   \State $\ybm = P\mu$ \label{p:p}
   \State $v_{\mu, \, \ell} = \sum_i C_{i \ell} - (u_{\mu, \, \ell} - \sm \ybml) \; \; \forall \ell$ \label{p:psi1fixed}
   \For {$k^\prime \leftarrow 1$, $N_y$}
   \Comment{Biconvex alternation loop}
      \State $b_{i \ell} = \ulil + \sl \yblil - \exp(-\yml)
             \; \; \forall i,\ell$
      \State $\ylil = \left( b_{i \ell} + \sqrt{b^2_{i \ell} + 4 \sl C_{i \ell}} \right)/(2\sl)
             \; \; \forall i,\ell$  \label{p:ylupdate}

      \State $\yll = \sum_i \ylil \; \; \forall \ell$

      \State $ \ypml = 0 \;\; \forall \ell$
      \Comment{Initialize Newton iteration}
      \For {$k^{\prime \prime} \leftarrow 1$, $N_\text{newt}$} 
      \Comment{Loop for solving Eq.~(\ref{yoptmu})}
          \State $ \psi^{(1)}_\ell = -\exp (-\ypml) \cdot \yll +\sm \ypml + v_{\mu , \, \ell}$ \label{p:psi1}
          \State $ \psi^{(2)}_\ell =  \exp (-\ypml) \cdot \yll +\sm \;\; \forall \ell$ \label{p:psi2}
          \State $ \ypml \leftarrow \ypml  - \psi^{(1)}_\ell \cdot \left( \psi^{(2)}_\ell\right)^{-1} \;\; \forall \ell$ \label{p:newt}
          \State $ \ypml \leftarrow \pos (\ypml) \;\; \forall \ell$  \label{p:pos}
          \Comment{Nonneg.,  Eq.~(\ref{yoptmu})} 

      \EndFor
      \State $ \yml = \ypml \;\; \forall \ell$

   \EndFor
   \State $\ul \leftarrow  \ul + \sl (\ybl  - \yl ) $ \label{p:ulupdate}
   \State $\um \leftarrow  \um + \sm (\ybm  - \ym ) $ \label{p:umupdate}
\EndFor
\end{algorithmic}
\hrulefill
\caption{ADMM pseudocode for SAA estimation with biconvex optimization. Variables
$\lambda$, $\mu$, $y_\lambda$, $y_\mu$, $\ybl$, $\ybm$, $\ul$, and $\um$ are initialized
to zero. Step size ratio parameters $\rho_\lambda$ and $\rho_\mu$ are chosen, and step size
parameters $\sl$, $\sm$, $\tl$, and $\tm$ are determined according to Eqs.~(\ref{stepl}) and (\ref{stepm}).
}
\label{alg:ADMM-SAA}
\end{algorithm}

\subsubsection*{ADMM pseudocode for SAA estimation}
The $x$-, $y$-, and $u$-update equations are assembled into a complete pseudocode given in
Algorithm~\ref{alg:ADMM-SAA}. The expensive projection and back-projection computations are collected in
as few lines as possible, and their
results stored, to avoid unnecessary repetition of these burdensome operations.
The simplex projection at line~\ref{p:lupdate} is the optimization problem defined
in Eq.~(\ref{lamopt2}); efficient computer code for implementing this projection is available
from Duchi {\it et al.} \cite{duchi2008efficient}.
The first derivative computation from Eq.~(\ref{psi1}) is performed at lines~\ref{p:psi1fixed} and \ref{p:psi1},
where line~\ref{p:psi1fixed} collects all terms that are not dependent on $\yl$ or $\ypm$.
The function $\pos (\cdot)$ in
lines~\ref{p:mupdate} and
\ref{p:pos} returns the argument if it is non-negative, otherwise it returns zero.
For the results presented in this work, we only consider zero initialization for all of the algorithm variables.
The choice of step size ratios $\rho_\lambda$ and $\rho_\mu$ will impact the convergence rate of the algorithm, and these
parameters must be tuned for optimal performance.

\subsection{ADMM for TV-constrained SAA in TOF-PET}
\label{sec:tvsaa}

The proposed ADMM framework for solving SAA estimation in TOF-PET allows for great flexibility in
imposing convex constraints in the imaging optimization problem. Accordingly, we augment the total
annihilation count and 
nonnegativity constraints in Eq.~(\ref{opt2}) with additional total variation
constraints on the activity and attenuation maps
\begin{multline}
\label{opt3}
\lambda,\mu = \argmin_{\lambda,\mu} \bigl\{ l(\lambda,\mu) \; | \; \;
\|\lambda\|_\text{TV} \le \gl, \; \; \|\mu\|_\text{TV} \le \gm, \bigr. \\
\bigl.
\mathbf{1}^\top \lambda = N_\text{total}, \; \; \lambda \ge 0, \; \; \mu \ge 0 \bigr\},
\end{multline}
where $\| \cdot \|_\text{TV}$ is the isotropic TV seminorm; $\gl$ and $\gm$ are the TV constraint values
for the activity and attenuation maps, respectively. The additional TV constraints exploit gradient
sparsity in in both the activity and attenuation that potentially improves accurate estimation of their
corresponding images. 

Because the novel aspect of this work is the treatment of the biconvex log-likelihood term,
which is explained in detail in Sec.~\ref{sec:ADMM-SAA}, the ADMM instance for this optimization problem
is covered in the Appendix.
The ADMM algorithm for TV-constrained SAA estimation (ADMM-TVSAA) is also
designed so that it makes use of the same step size ratio parameters as discussed for Algorithm~\ref{alg:ADMM-SAA}.
Because of the additional constraints, the TV constraint values $\gl$ and $\gm$ become additional parameters
of the algorithm.

\subsection{Step size scaling of ADMM for SAA}
When tuning the step size parameters $\rho_\lambda$ and $\rho_\mu$ in ADMM-SAA or ADMM-TVSAA, it is important
to account for the fact that the the optimal settings for maximum algorithm efficiency change with scaling
of the coincidence data $C$. A practical consequence is that optimally efficient $\rho$-values would depend
on, for example, collection time of the TOF-PET system. In the following results presented in Sec.~\ref{sec:results},
this issue is addressed by normalizing the data $C$ with the factor $\text{size}(C)/\|C\|_2$. The choice
of data normalization is arbitrary because of the following scaling relationship. If we replace the coincidence
data $C$ by $aC$, the same ADMM iterates, up to a scaling,
can be obtained by adjusting the constraint and step size parameters
as follows:
\begin{multline*}
N_\text{total} \rightarrow a N_\text{total}, \;\; \gl \rightarrow a \gl, \\
\sigma_\lambda \rightarrow \sigma_\lambda/a, \;\; \tau_\lambda \rightarrow a \tau_\lambda, \;\;
\sigma_\mu \rightarrow a \sigma_\mu, \;\; \tau_\mu \rightarrow \tau_\mu/a.
\end{multline*}
With this scaling of the algorithm parameters, the algorithm variables transform as follows:
\begin{align*}
\lambda & \rightarrow a \lambda, \;\; \yl \rightarrow a \yl, \;\; \ul \rightarrow \ul, \\
\mu & \rightarrow  \mu, \;\; \ym \rightarrow \ym, \;\; \um \rightarrow a \um.
\end{align*}
The scaling transformations can be verified by making the appropriate substitutions into
Eqs.~(\ref{nadmm3}), (\ref{lamopt}), (\ref{muopt}), (\ref{yoptlam}), and (\ref{yoptmu}), or
into the update equations of Algorithm 1.

\subsection{Huber-penalized MLAA}
\label{sec:hup}
Use of ADMM allows for nonsmooth terms in the optimization such as use of the TV-norms and complex constraints
on the activity and attenuation. The closest comparison from the literature involves a smooth objective function
with Huber penalties on the activity and attenuation.
Accordingly comparison results are
obtained with the penalized-MLAA (P-MLAA) algorithm presented in Mehranian {\it et al.} \cite{mehranian2017mr}.
The P-MLAA algorithm implemented here addresses the following optimization problem Huber penalties
\begin{multline}
\label{pmlaa}
\lambda,\mu = \argmin_{\lambda,\mu} \bigl\{ l(\lambda,\mu) +\beta H_\delta(\lambda) + \gamma H_\delta(\mu)  \; | \; \;
\bigr. \\
\bigl.
\mathbf{1}^\top \lambda = N_\text{total}, \; \; \lambda \ge 0, \; \; \mu \ge 0 \bigr\},
\end{multline}
where $H_\delta(\cdot)$ is the Huber penalty with smoothing parameter $\delta,$ see for example Nuyts {\it et al.} \cite{nuyts1999simultaneous}
for the definition of this penalty function. The P-MLAA algorithm replaces Eq.~(\ref{mlaa2}) with the ``one step late'' update
equation developed by Green \cite{green1990bayesian} and introduces a Gibbs prior into Eq.~(\ref{mlaa4}) in the manner developed
in Nuyts {\it et al.} \cite{nuyts1999simultaneous}. 

\begin{figure}[!t]
\centerline{\includegraphics[width=0.4\columnwidth]{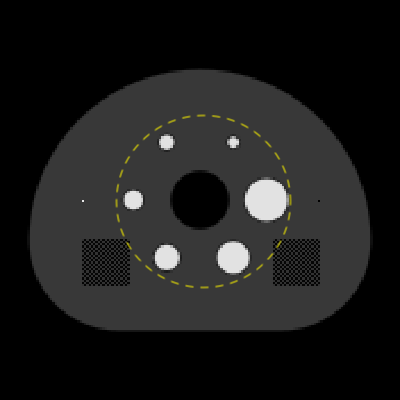}
\includegraphics[width=0.4\columnwidth]{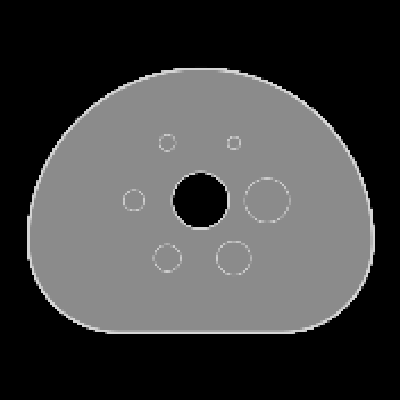}}
\caption{(left) Slice number 40 from the University of Washington Digital Reference Object:
activity image in arbitrary units, and (right) attenuation map displayed in the gray
scale window $[0.075,0.115]$ cm$^{-1}$. The dashed circle in the activity image indicates
the activity distribution used for the investigation of SAA with interior data.}
\label{fig:phantom}
\end{figure}

\section{Results with a 2D TOF-PET simulation}
\label{sec:results}

The results demonstrating the ADMM-SAA algorithm are all derived from a 2D 
simulation using the digital reference object (DRO) shown in Fig.~\ref{fig:phantom} \cite{pierce2015digital}.
This digital phantom is cropped to 176x176 image array with physical dimension 30x30 cm$^2$.
The LORs are arranged in a 2D parallel-beam geometry with 176 views covering a $\pi$ radian arc,
and 176 parallel rays being measured per view with a spacing of 0.17 cm (30/176 $\approx$ 0.17).
The TOF FWHM is taken to be 4.5 cm, which corresponds to a timing resolution of approximately 300 picoseconds.
The spacing between TOF window samples is 2.25 cm, and a total of 17 TOF samples are taken per LOR.
For the image reconstruction, both the attenuation and activity images are represented on a 176x176 grid.
The purpose of the presented results is to demonstrate usage of the ADMM-SAA algorithm and the impact
of the TV constraints on the reconstruction of the activity and attenuation.

For the following results, the biconvex alternation loop at line~9 of Algorithm 1
is run for $N_y=100$ iterations, and the Newton solver at line~14 is run for $N_\text{newt}=10$
iterations. With both of these loop settings, Eq.~(\ref{gdef}) is solved accurately in a numerical sense.
Even with the inner loops being executed with such high iteration numbers, the efficiency of the whole
biconvex alternation loop is still high, because all of the computations separate across the vector components.
The computational effort for the biconvex alternation loop is $O(N_y \cdot N_\text{TOF} \cdot N_\text{views} \cdot \sqrt{N_\text{pix}})$
(the Newton loop does not increase the order of this loop because it involves the attenuation sinogram only),
and by comparison, computing TOF projection, $T\lambda$, is  $O(N_\text{TOF} \cdot N_\text{views} \cdot N_\text{pix})$,
where $N_\text{pix}$ is the total number of pixels and $N_\text{views}$ is the number of projection angles.
For the small 176x176
images of this study the biconvex loop is of the same order as TOF projection
because $N_y \approx \sqrt{N_\text{pix}}$, but as the
data and image size increase, TOF projection becomes the more burdensome computation.
It is also possible, in practice, to reduce $N_y$ and $N_\text{newt}$ and work with inexact solution
of Eq.~(\ref{gdef}) but we do not investigate this option in this work.

\subsection{SAA from noiseless data}
\label{sec:noiseless}

\begin{figure}[!t]
\centerline{
\includegraphics[width=0.49\columnwidth]{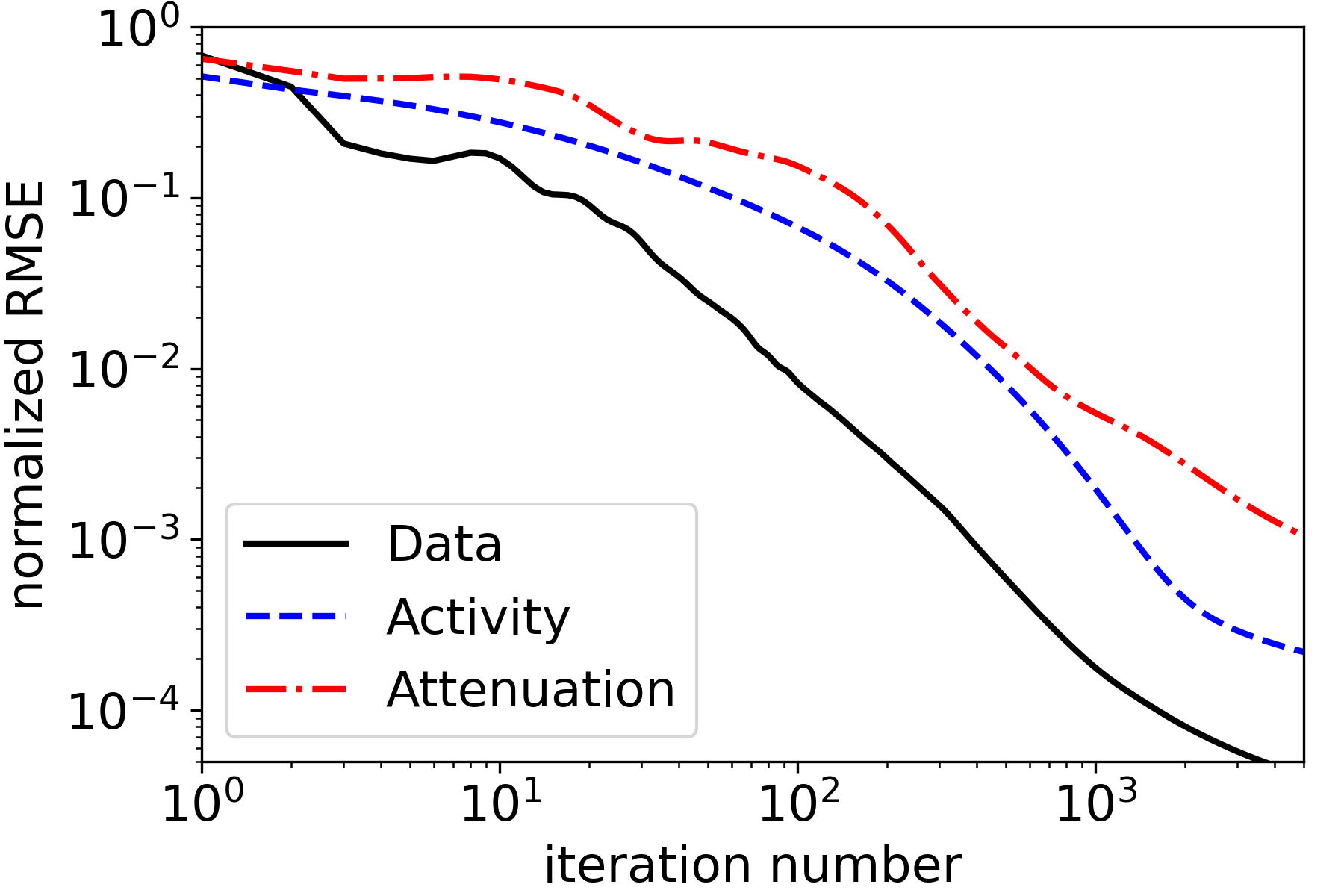}
\includegraphics[width=0.49\columnwidth]{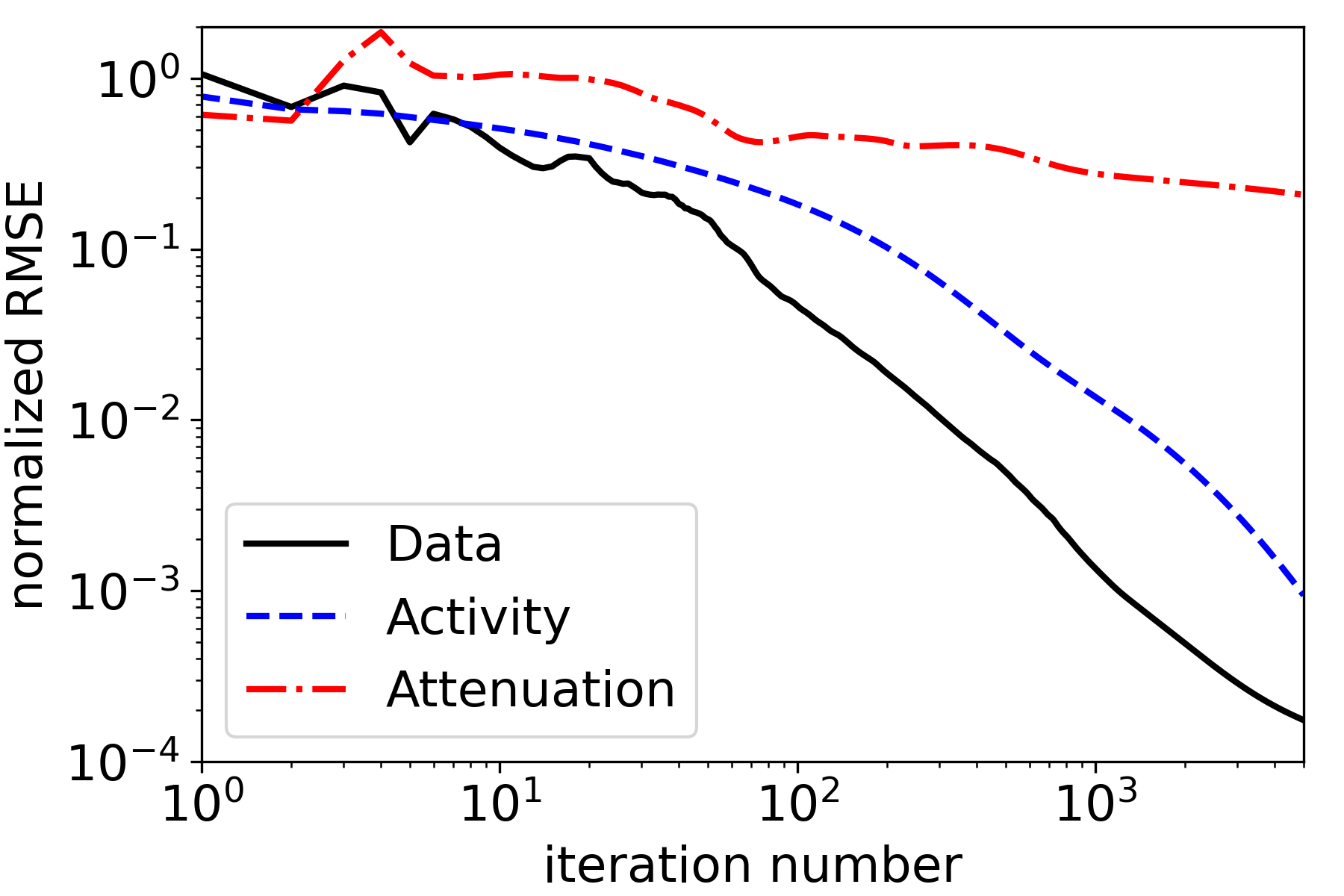}
}
\caption{Convergence of ADMM-TVSAA with noiseless TOF-PET data for the case of the full activity
distribution (left) and the interior activity distribution (right). The data RMSE is normalized to the mean value
of the TOF-PET data, and the activity/attenuation RMSEs are normalized to the mean values of their respective images.}
\label{fig:icrmseTV}
\end{figure}

Image reconstruction is performed on noiseless data using the mean counts as the measured data, and
ADMM-TVSAA is employed to study the effectiveness
for solving the associated inverse problem
for two situations: (1) the full activity distribution from the DRO phantom is used as the test object,
and (2) the activity distribution is truncated at the yellow, dashed circle in Fig.~\ref{fig:phantom}.
The second case is a more challenging inverse problem for SAA because recovery of the full attenuation
map is complicated by the fact that only LORs that intersect the non-zero activity provide useful data,
and accordingly recovery of the attenuation map has a similar degree of difficulty as the interior problem
in tomography.

The convergence results for 5000 iterations of ADMM-TVSAA on the phantom with the full activity distribution
and the truncated (or interior) activity distribution
are shown in Fig.~\ref{fig:icrmseTV}.
In both cases, the normalized data RMSE and activity RMSE is observed to steadily converge to zero although the
activity RMSE is noted to converge more slowly than the data RMSE. Recovering the true attenuation is clearly more
challenging; for the full activity distribution case the attenuation RMSE exhibits convergent behavior
all the way to the last computed iteration,
but for the truncated activity case the global
attenuation RMSE does not demonstrate convergent behavior. For truncated
activity, however, the attenuation factor derived from the inaccurate attenuation map must have high
degree of accuracy since the activity distribution is nicely recovered.

\begin{figure}[!t]
\centerline{
\includegraphics[width=0.49\columnwidth]{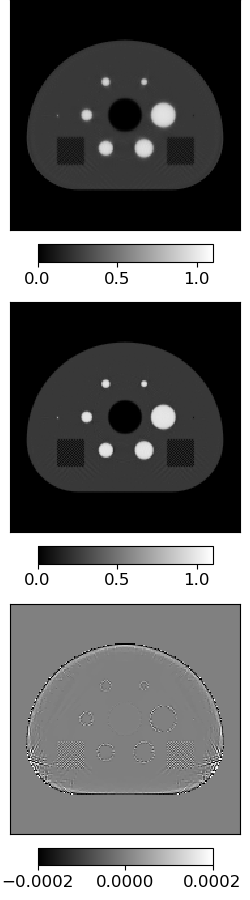}
\includegraphics[width=0.49\columnwidth]{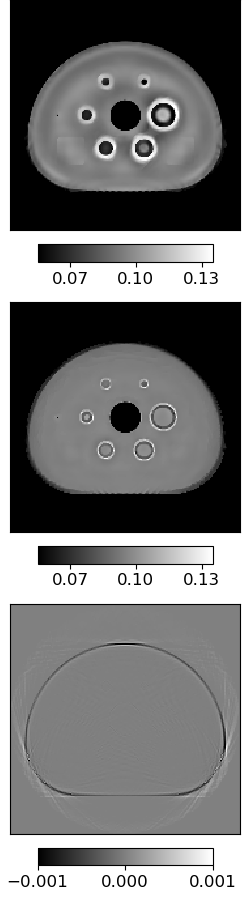}
}
\caption{Reconstructed activity (left column) and attenuation (right column) images from noiseless data with ADMM-TVSAA
at 50 (top row), 100 (middle row), and 5000 (bottom row) iterations. Because the result at 5000 iterations is visually
indistinguishable from the test phantom the difference from the ground truth is displayed in the bottom row.
The activity distribution is normalized to 1.0 for the maximum value.}
\label{fig:imageseriesTV}
\end{figure}

\begin{figure}[!t]
\centerline{
\includegraphics[width=0.49\columnwidth]{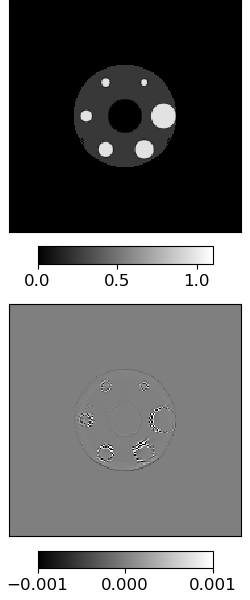}
\includegraphics[width=0.49\columnwidth]{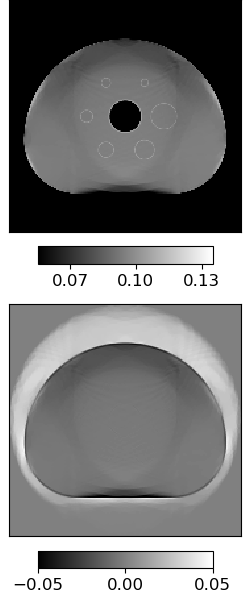}
}
\caption{For the results with the interior activity distribution we only show iteration 5000
for the activty (left) and attenuation (right). The actual activity/attenuation images are shown
in the top row, and the difference from ground truth is shown in the bottom row.
}
\label{fig:interiorseriesTV}
\end{figure}

\begin{figure}[!t]
\centerline{~~~~~~Activity~~~~~~~~~~~~~~~~~~~~~~~~~Attenuation}
\centerline{\includegraphics[width=0.49\columnwidth]{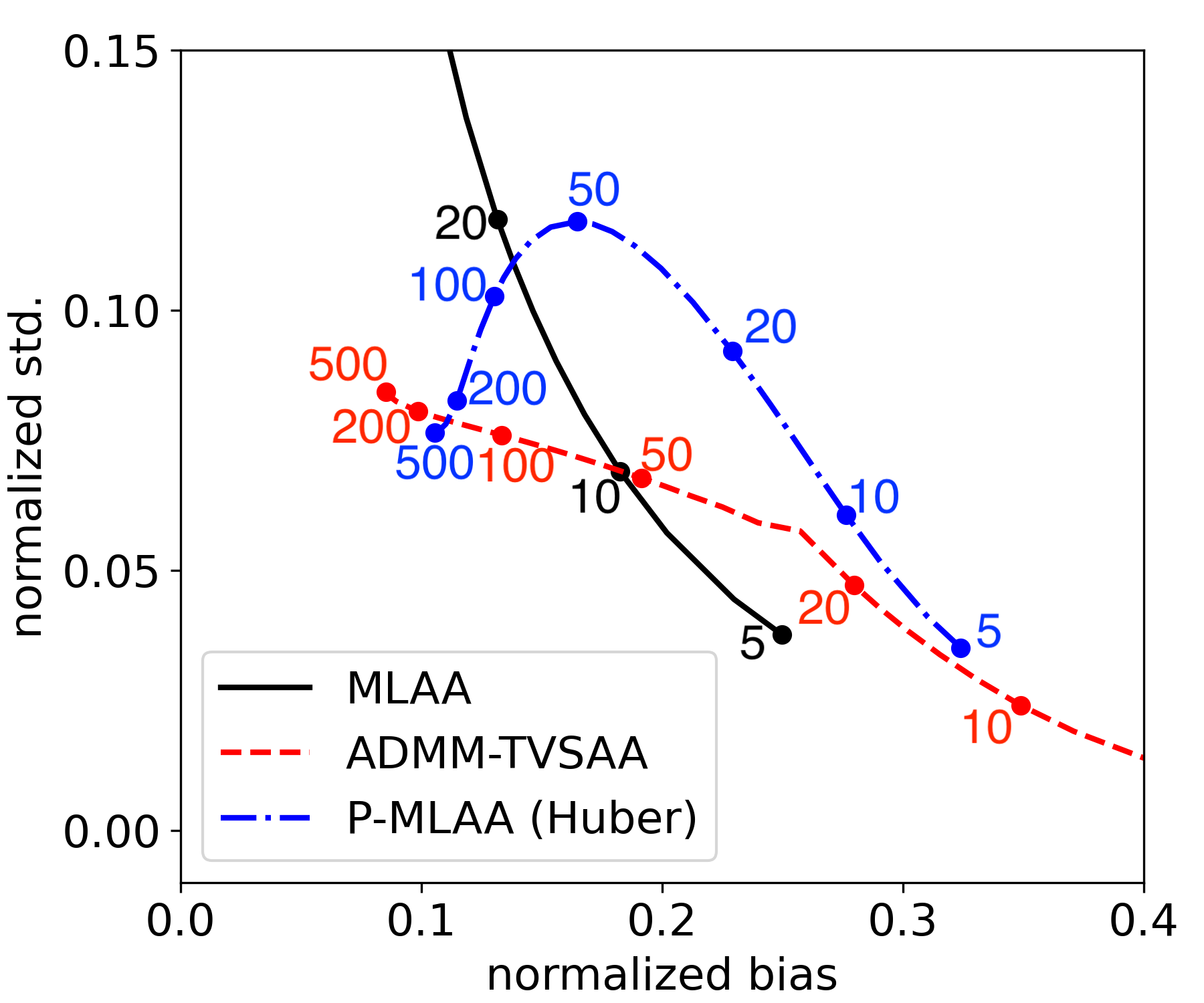}
\includegraphics[width=0.49\columnwidth]{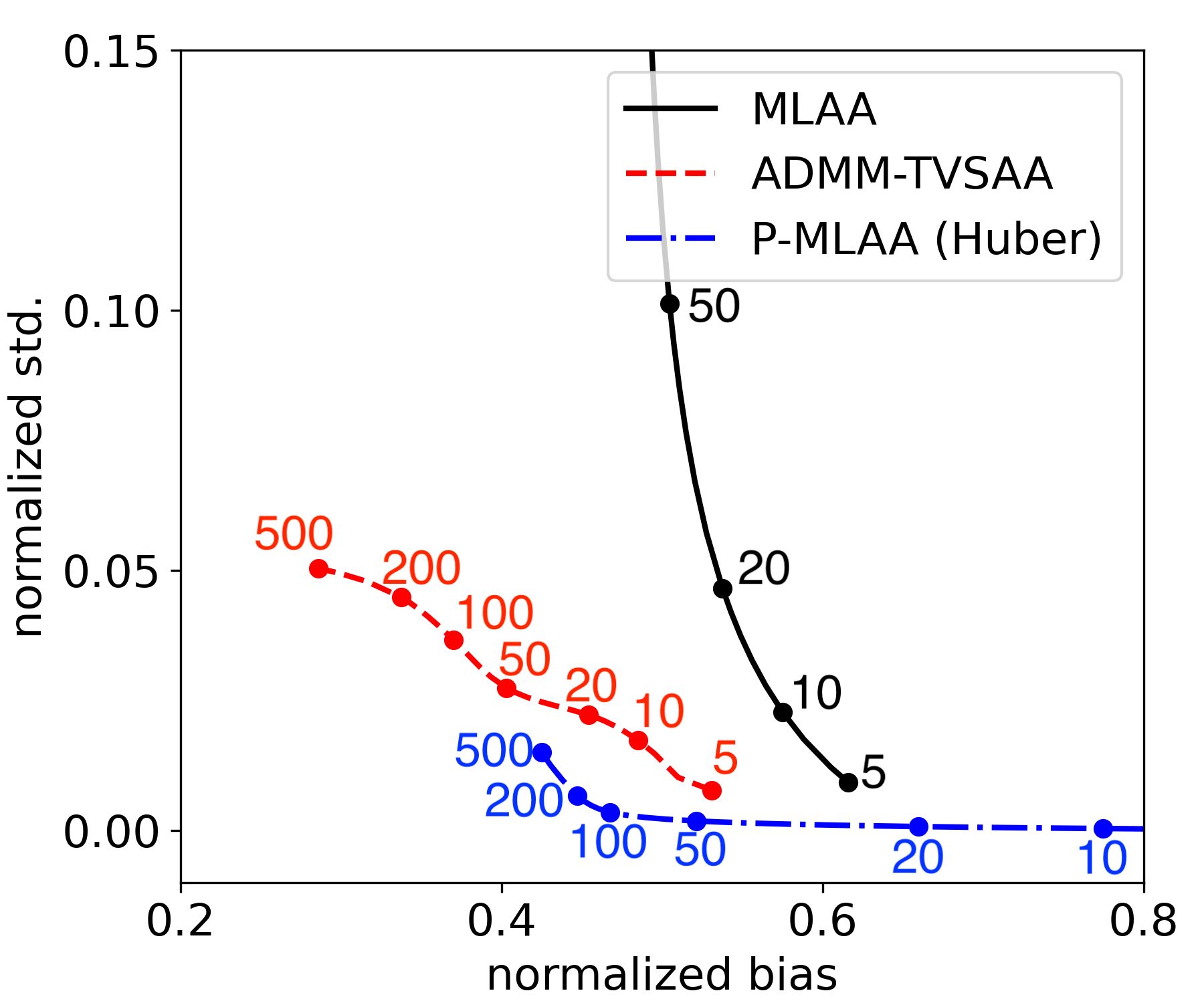}}
\caption{Normalized standard deviation versus normalized bias of the activity (left) and
attenuation (right) images as a function of iteration number
computed empirically from 100 noise realizations for MLAA, MLAA with Huber penalties, and TV-constrained SAA.
Normalization of bias and standard deviation is achieved by dividing by the mean value of the corresponding
ground truth image.
The labeled dots indicate the iteration numbers for the respective algorithm curves. For TV-constrained
ADMM-SAA, curves are shown for activity and attenuation
TV constraints set to $\gamma_\lambda = 1.0$ and $\gamma_\mu = 1.0$, respectively, where 
the constraint values are given as a fraction of the ground truth
TV values. The Huber penalty parameters for P-MLAA
are set so that the resulting activity and attenuation images
have nearly the same TV values as the ground truth after 500 iterations.}
\label{fig:bias_std}
\end{figure}

A series of image estimates are shown for the full activity in Fig.~\ref{fig:imageseriesTV}
and only the 5000th iteration results are shown for truncated activity in Fig.~\ref{fig:interiorseriesTV}.
The difference images at 5000 iterations
shows accurate reconstruction of both activity and attenuation for the full distribution case. The attenuation
difference
in the bottom panels reveals error at the 1\% level for the attenuation map near the edges of the phantom.
The panels at the earlier iterations provide a sense of the convergence to the solution; specifically the activity
images at 50 or 100 iterations closely resemble the result at 5000 iterations. The attenuation map clearly converges
more slowly.
For the case of truncated activity shown in Fig.~\ref{fig:interiorseriesTV}, the activity image
is accurately recovered but it is clear that
the attenuation map is not completely recovered.
Interestingly, ADMM-TVSAA does seem to be able to recover the support of the attenuation map even if there is
substantial error in the outer portions of the image. Moreover, the central portion of the attenuation image
in the location where the activity is non-zero does appear to be reconstructed accurately.

\subsection{SAA from noisy data}
\label{sec:noisy}

The next set of studies focus on SAA with noisy data
and only the case of the full activity distribution is considered.
Noise realizations are obtained by scaling the mean
TOF-PET data so that the total number of measured coincidences is 
4$\times$10$^6$; the realization is then obtained
by selecting a number of detected coincidences for each time-window sample and LOR, drawn from a Poisson distribution.
In demonstrating the use of ADMM-TVSAA, two forms of MLAA provide reference algorithms. The
MLAA algorithm described in Sec.~\ref{sec:mlaa} is one of the references, where early stopping provides
regularization. The P-MLAA algorithm with Huber penalties provides the other reference where the parameters
are chosen in such a way that it could conceivably yield similar results as ADMM-TVSAA after 500 iterations.
All three algorithms enforce the total annihilation count and non-negativity constraints. The smoothing parameter
for P-MLAA's Huber functions are both chosen to be 0.1\% of the phantom maximum value,
which is much less than the contrast of structures in either
the attenuation or activity maps. In this way the Huber penalties approximate the TV-norm accurately. The penalty
parameters are tuned so that the phantom TV values are achieved at 500 iterations of P-MLAA.

\begin{figure}[!t]
\centerline{
\includegraphics[width=0.9\columnwidth]{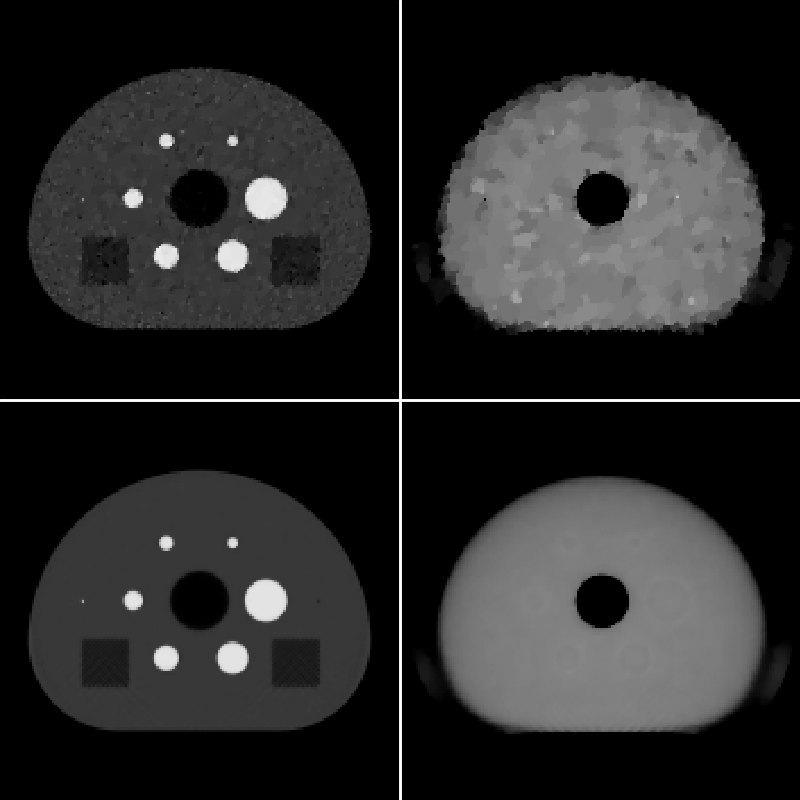}
}
\caption{Reconstructed activity (left column) and attenuation (right column) images from noisy data with ADMM-TVSAA.
The top row shows a reconstructed set of images from a single noise realization at 500 iterations, and the 
bottom row shows the corresponding mean over 100 noise realizations.
With the activity distribution normalized to 1.0 for the maximum value, the gray scale for the activity images
is $[0,1.1]$, and the gray scale for the attenuation is $[0.055,0.135]$.}
\label{fig:noisyimagesTV}
\end{figure}

For a quantitative bias-variance analysis of the activity and attenuation, MLAA, P-MLAA, and ADMM-TVSAA
are used to perform SAA on an ensemble
of 100 noise realizations of TOF-PET data. The mean and pixel standard deviation 
are computed and plotted in Fig.~\ref{fig:bias_std} as a function of iteration number.
The use of the TV-constraints, allows ADMM-TVSAA to achieve activity estimates with low bias and variance
as compared with basic maximum-likelihood estimation as implemented with MLAA.
Use of explicit Huber penalties with P-MLAA also yields images at 500 iterations
that have low bias and variance with respect to MLAA; although the paths in the bias-variance plot
for ADMM-TVSAA and P-MLAA are quite different as a function of iteration number. For the particular
parameter settings chosen, the ADMM-TVSAA
algorithm achieves slightly lower global bias in the activity image with slightly larger variance
as compared to P-MLAA.
In the activity bias-variance curves, the proximity of the points at 200 and 500 iterations
is an indication that the respective ADMM-TVSAA and P-MLAA are near convergence.

The bias-variance curves for the attenuation image reveal a much different behavior
than that of the activity image. Both ADMM-TVSAA and P-MLAA substantially improve
on the use of MLAA without explicit penalty terms. The variance of the ADMM-TVSAA
result is larger than that of P-MLAA, but ADMM-TVSAA achieves a lower bias.
It is also clear that 500 iterations is not sufficient to achieve a converged attenuation
map because the there is still quite some separation between the points at 200 and
500 iterations for ADMM-TVSAA and P-MLAA. This difference in the rate of convergence of the
activity and attenuation images was also seen in the noiseless results shown in Fig.~\ref{fig:icrmseTV}.
That there can be such a difference in the convergence rate of the two images is due
to the fact that the attenuation image only impacts the activity image through the attenuation
factor.

\begin{figure}[!t]
\centerline{
\includegraphics[width=0.9\columnwidth]{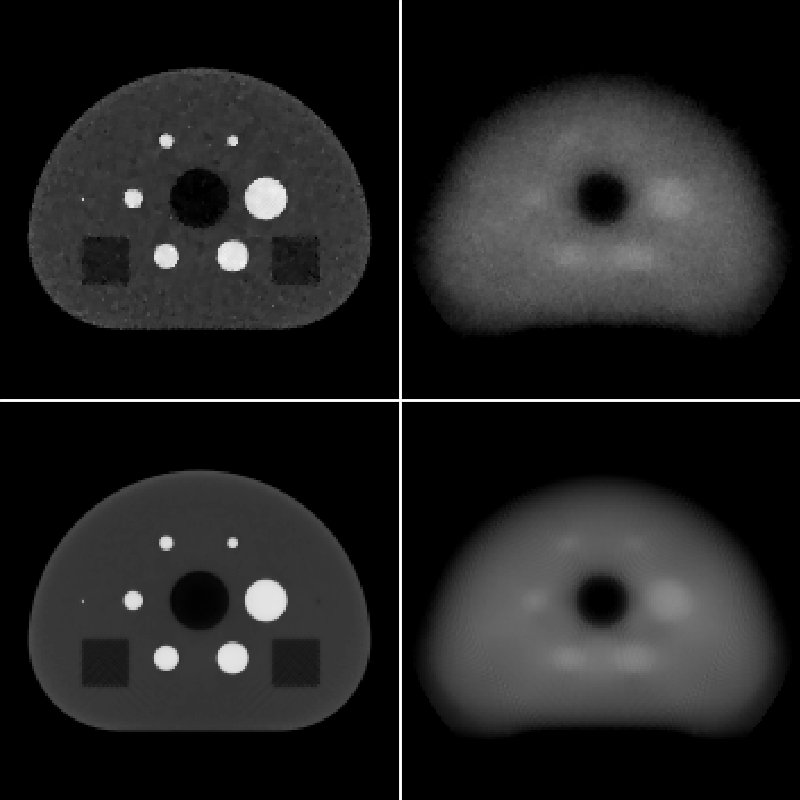}
}
\caption{Same as Fig.~\ref{fig:noisyimagesTV} except that the Huber-regularized MLAA is used
to generate the image iterates.}
\label{fig:noisyimagesHU}
\end{figure}

\begin{figure}[!t]
\centerline{
\includegraphics[width=0.49\columnwidth]{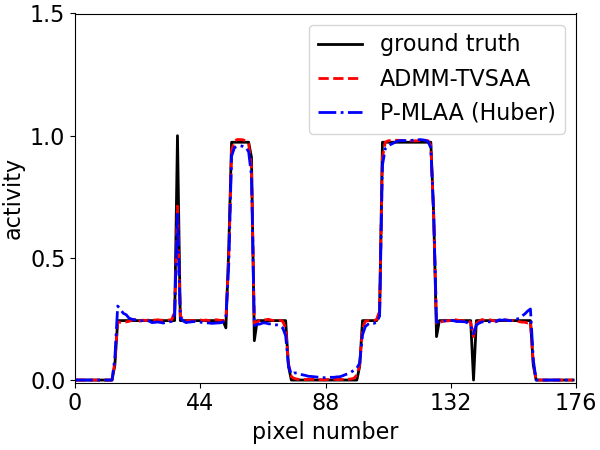}
\includegraphics[width=0.49\columnwidth]{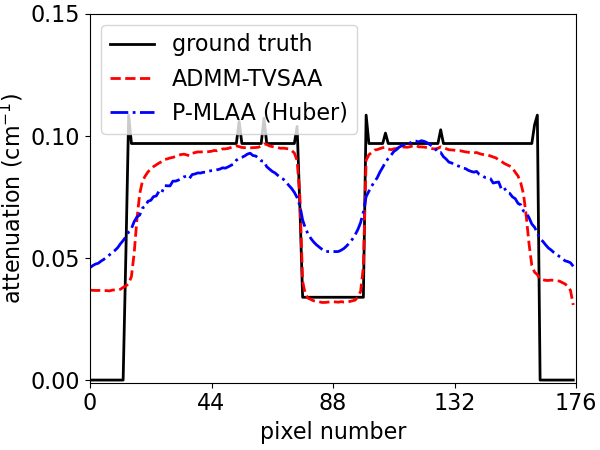}
}
\caption{Line profiles through the middle row of pixels comparing the ground truth
with the mean reconstructed activity (left) and attenuation (right) images generated
at 500 iterations by ADMM-TVSAA and P-MLAA.}
\label{fig:noisyprofiles}
\end{figure}

The activity and attenuation images for the present noise studies
are shown in
Figs.~\ref{fig:noisyimagesTV} and \ref{fig:noisyimagesHU} for ADMM-TVSAA and P-MLAA,
respectively. The top row of each figure shows images at 500 iterations for a single
noise realization, and the bottom row shows the mean over 100 noise realizations, which
reveals the spatial dependence of the image bias. The most notable
difference between the two algorithms is in the attenuation images. The ADMM-TVSAA
algorithm achieves
an attenuation distribution that has the correct support and gray level even if it suffers from
noticeable noise artifacts. The result for P-MLAA, however, shows significant bias in the attenuation
images and features from the activity distribution clearly bleed through to the attenuation images.
The error in the attenuation images, however, may not be critical if the attenuation factors are recovered,
and this appears to be the case for both algorithms as the activity images at 500 iterations are
accurate. 
The activity for P-MLAA shows slightly more bias, relative to ADMM-TVSAA, at the phantom border.

The line profiles in Fig.~\ref{fig:noisyprofiles} reveal the bias more quantitatively.
For ADMM-TVSAA, the mean activity follows the line profile of the ground truth quite closely
except for the fact that the sharp features are slightly rounded. The mean attenuation
is somewhat accurate except for near the border of the object support where there is significant
blurring of the edge. For P-MLAA, the mean activity has a slight increase at the phantom border
and a slight filling in of the central cold spot. The attenuation profile for P-MLAA reflects
the significant bias that was seen in Fig.~\ref{fig:noisyimagesHU}.

\begin{figure}[!t]
\centerline{
\includegraphics[width=0.49\columnwidth]{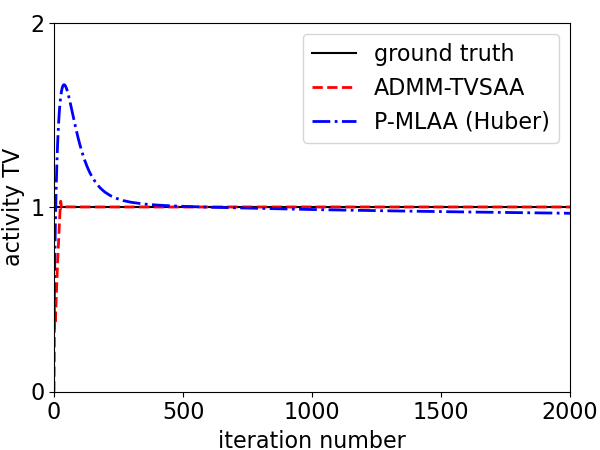}
\includegraphics[width=0.49\columnwidth]{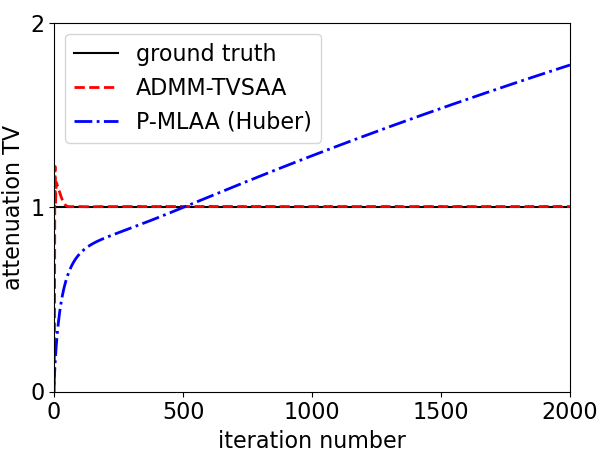}
}
\caption{Evolution of activity (left) and attenuation (right) image TV as a function
of iteration number for 2000 iterations of ADMM-TVSAA and P-MLAA. The TV values are normalized
so that the ground truth phantom values are 1.0.}
\label{fig:tvevol}
\end{figure}

\section{Discussion}
\label{sec:discussion}
The constrained likelihood model for SAA in TOF-PET
enables novel numerical investigation into the SAA inverse problem. 
In particular, for this work, TV-constraints that exploit gradient sparsity
are investigated for simultaneous recovery of activity and attenuation images.
From the formulation presented in Sec.~\ref{sec:tvsaa}, given knowledge
of the total annihilation count and TV values of the activity and attenuation,
the question is whether or not it is possible to recover the underlying
activity and attenuation distributions. For the presented noiseless study where the activity
and attenuation have the same support, it appears that both distributions can be
accurately recovered. For the case where the activity distribution is interior
to the attenuation, the former is recovered accurately while the latter is recovered only
within the support of the activity distribution. The study with data noise
provides some sense of the stability of the SAA inverse problem when using the
TV-constrained likelihood model.


We point out that carrying out inverse problems studies with the penalized likelihood
approach shown in Sec.~\ref{sec:hup} is much more difficult.
Aside from the difference between TV and Huber regularization terms, penalized
likelihood needs to be investigated in the limit where the penalty parameters
approach zero in order to provide the equivalent solution to constrained likelihood
when the data are noiseless. 

Even with noisy data, it is difficult to use
the penalized-likelihood approach for achieving the same solution as the constrained-likelihood
model. This is illustrated in the evolution of the activity and attenuation image TV shown in
Fig.~\ref{fig:tvevol}. These results correspond to the noise study that yielded the
single noise realization images in Figs.~\ref{fig:noisyimagesTV} and \ref{fig:noisyimagesHU}
except that the iteration number is extended to 2000. For ADMM-TVSAA, the image TVs
reach their constraint values quickly and maintain these values. For P-MLAA, the desired
constraint values are achieved at 500 iterations because a two-dimensional penalty
parameter search was performed to achieve this goal. Iterating further with P-MLAA
causes the image TV to change. This change is not large for the activity because
this image is nearly converged, but for the attenuation the change in TV is rather large.
In order to come closer to the desired constraint values with P-MLAA, the penalty parameter
search would need to be performed at larger iteration numbers, increasing the computational
burden.

So far this discussion has focused on algorithms for investigating the TOF-PET SAA inverse problem,
which can be quite different than their use for image reconstruction in real data scenarios.
For an actual scan, ground truth is not available and mathematical accuracy of the solution
may not correlate well with the imaging task of interest. Parameters of ADMM-TVSAA or P-MLAA
need to be optimized on a metric reflecting performance on this imaging task. Also, because
accurate solution is not necessarily the goal, early stopping is almost always used in practical
applications of iterative image reconstruction. In this sense, iteration number becomes a parameter
to tune and
the practical difference between these two algorithms is the different paths they take in
approaching their respective solutions.
We also point out that we have shown results for P-MLAA using Green's one step
late algorithm, and use of other solvers such as ADMM with SQS \cite{chun2016joint} may lead to
improved performance of P-MLAA.

The presented study made use of a relatively simple test object that may favor the use of TV constraints.
For realistic activity distributions, such constraints can still be employed effectively as is
shown in Refs.~\cite{zhang2016investigation} and \cite{zhang2018optimization}, where human PET data
is reconstructed using such constraints.

\section{Conclusion}
\label{sec:conclusion}

In this work, an ADMM framework is developed that can be applied to nonsmooth and nonconvex optimization 
problems that arise in imaging. The particular form of nonconvexity addressed is when the optimization problem
has a biconvex structure. The imaging problem posed by simultaneous estimation of the activity and attenuation (SAA)
in time-of-flight positron emission tomography (TOF-PET) has such a structure. Using this ADMM framework,
a limited study on the impact of total variation (TV) constraints on the activity and attenuation for the SAA
problem is presented. The use of both of these constraints is seen to help stabilize the SAA inverse problem.
The shown results are intended to demonstrate the ADMM-TVSAA algorithm and its
potential in solving the SAA inverse problem with the use
of TV constraints on the activity and attenuation.
While we have shown results only for ADMM-TVSAA, the ADMM framework is easily extended to include other
nonsmooth, convex terms.
Further study varying the test phantom and TOF-PET setup are needed to obtain a more comprehensive picture
of the SAA inverse problem.

\section*{Data Availability}
The implementation of the algorithms, which are presented in this article, and the code, which generates the figures,
are available at:\\
\url{https://github.com/zhimeir/saa_admm_paper}.

\appendix

\subsection*{ADMM for TVSAA in TOF-PET}

In this Appendix, the details for the instantiation of ADMM for TV-constraineed SAA estimation (ADMM-TVSAA)
is explained. The optimization problem of interest is Eq.~(36) from the main text, which we
restate here
\begin{multline}
\tag{A1}
\lambda,\mu = \argmin_{\lambda,\mu} \bigl\{ l(\lambda,\mu) \; | \; \;
\|D \lambda\|_1 \le \gl, \; \; \|D \mu\|_1 \le \gm, \bigr. \\
\bigl.
\mathbf{1}^\top \lambda = N_\text{total}, \; \; \lambda \ge 0, \; \; \mu \ge 0 \bigr\},
\end{multline}
where $D$ is the discretization of the spatial gradient operator, and $\|D x\|_1$ is the anisotropic TV of $x$.
To map the optimization problem onto the generic ADMM optimization in Eq.~(14),
the primal, splitting, and dual variables $x$, $y$, and $u$, are respectively assigned as
\begin{equation*}
x = \left(
\begin{array}{c}
\lambda \\
\mu
\end{array}
\right),  \; \;
y = \left(
\begin{array}{c}
y_\lambda \\
z_\lambda\\
y_\mu\\
z_\mu
\end{array}
\right), \; \;
u = \left(
\begin{array}{c}
u_\lambda \\
v_\lambda\\
u_\mu\\
v_\mu
\end{array}
\right),
\end{equation*}
The linear system $A$ is assigned as
\begin{equation*}
A = \left(
\begin{array}{cc}
T & 0\\
\nu_\lambda D & 0\\
0 & P\\
0 & \nu_\mu D
\end{array}
\right),
\end{equation*}
where 
\begin{equation}
\tag{A2}
\nu_\lambda = \|T\|_2/\|D\|_2, \; \; \;\nu_\mu = \|P\|_2/\|D\|_2,
\end{equation}
are constants that normalize the gradient matrices to the projection matrices.

As with the SAA problem,
the convex function $f$ is used to represent the constraint on the total number of annihilations
by setting
\begin{equation*}
f(\lambda,\mu) = \delta( \mathbf{1}^\top \lambda = N_\text{total}) + \delta (\lambda \ge 0) + \delta (\mu \ge 0).
\end{equation*}
The biconvex function $g$ accounts for the remaining terms in Eq.~(A1)
\begin{multline}
\tag{A3}
g(\yl,\ym) = L(\yl,\ym) + \\
\delta(\|z_\lambda\|_1 \le \nu_\lambda \gamma_\lambda) + \delta(\|z_\mu\|_1 \le \nu_\mu \gamma_\mu),
\end{multline}
where the biconvex function $L(\yl,\ym)$ is defined in Eq.~(23) and the TV constraint
values have also been scaled to reflect the normalization of $D$.

\subsubsection*{Parametrization of the step-sizes}
With the modified system matrix $A$, the metric from Eq.~(14) becomes
\begin{align*}
H_f =& \left(
\begin{array}{cc}
H_\lambda & 0\\
0 & H_\mu
\end{array}
\right), \\
H_\lambda =& \frac{I}{\tau_\lambda} - \sigma_\lambda( T^\top T + \nu^2_\lambda D^\top D), \\
H_\mu =& \frac{I}{\tau_\mu} - \sigma_\mu( P^\top P +\nu^2_\mu D^\top D),
\end{align*}
and the step-size parameters are chosen so that
\begin{equation*}
\sigma_\lambda \tau_\lambda = (\| T \|^2_2 + \nu^2_\lambda \|D\|^2_2)^{-1}, \; \;
\sigma_\mu \tau_\mu = (\| P \|^2_2 + \nu^2_\mu \|D\|^2_2)^{-1}.
\end{equation*}
As with ADMM-SAA
the step size ratios, $\rho_\lambda$ and $\rho_\mu$, need to be 
determined by a similar grid search to what is shown in Sec.~IIIA.
The grid search performed for ADMM-SAA provides a good initial starting point for the
step-size parameter search for
ADMM-TVSAA when the parameters $\nu_\lambda$ and $\nu_\mu$
are determined by Eq.~(A2). Accordingly, the step size parameters
for ADMM-TVSAA are
\begin{align*}
\sigma_\lambda  &= \rho_\lambda (\| T \|^2_2 + \nu^2_\lambda \|D\|^2_2)^{-1},\\
\tau_\lambda &= \rho_\lambda^{-1} (\| T \|^2_2 + \nu^2_\lambda \|D\|^2_2)^{-1}, \\
\sigma_\mu &= \rho_\mu(\| P \|^2_2 + \nu^2_\mu \|D\|^2_2)^{-1},\\
\tau_\mu &= \rho_\mu^{-1} (\| P \|^2_2 + \nu^2_\mu \|D\|^2_2)^{-1}.
\end{align*}

\subsubsection*{The $y$-update}
The $g$ function in Eq.~(A3) separates into a biconvex function in  $y_\lambda$ and  $y_\mu$,
and convex functions in $z_\lambda$ and $z_\mu$.
The biconvex terms are treated in exactly the same way as the ADMM-SAA presentation in Sec.~IIF.
Focusing on the last two convex terms in Eq.~(A3) yields the optimization problems
\begin{align}
&\zl = \tag{A4}\\
&\argmin_{\zpl} \Bigl\{
\delta(\| \zpl \|_1 \le \nl \gamma_\lambda)
- \vl^\top \zpl
+ \frac{\sl}{2} \| \zpl - \nl D\lambda \|^2
\Bigr\}, \notag
\end{align}
and
\begin{align}
&\zm = \tag{A5}\\
&\argmin_{\zpm} \Bigl\{
\delta(\| \zpm \|_1 \le \nm \gamma_\mu)
- \vm^\top \zpm
+ \frac{\sm}{2} \| \zpm - \nm D\mu\|^2
\Bigr\}. \notag
\end{align}
Because both of these problems are identical, we focus on Eq.~(A4).
The objective function consists of a quadratic function and an indicator function that enforces
the $\ell_1$ constraint on $\zpl$. Furthermore, the quadratic function has uniform curvature, i.e. a
Hessian matrix that is proportional to the identity matrix. For this special case, the solution to Eq.~(A4) 
is obtained in a two-step process that involves finding the
minimizer of the unconstrained quadratic function, then projecting the result to the closest $\zpl$
that satisfies the $\ell_1$ constraint.
The unconstrained minimizer is given by
\begin{equation*}
\zppl = \vl/\sl + \nl D \lambda,
\end{equation*}
and the solution to Eq.~(A4) becomes
\begin{equation*}
\zpl = \mathcal{P}_{L1(\nl \gamma_\lambda)}(\zppl), \; \; L1(r) = \{z \; | \; \| z \|_1 \le r\},
\end{equation*}
where $\mathcal{P}_{L1(\nl \gamma_\lambda)}(\cdot)$ denotes projection onto the
$\ell_1$-ball of ``radius'' $\nl \gamma_\lambda$. An efficient algorithm for performing
this projection is presented in Ref. \cite{duchi2008efficient}, or it can also be accomplished by
vector shrinkage and use of a root finding algorithm to determine the shrinkage parameter
to attain an $\ell_1$-norm of $\nl \gamma_\lambda$.
The update equations for $\zl$ and $\zm$ are 
\begin{align*}
\zl &\leftarrow \mathcal{P}_{L1(\nl \gamma_\lambda)}(\vl/\sl + \nl D \lambda),\\
\zm &\leftarrow  \mathcal{P}_{L1(\nm \gamma_\mu)}(\vm/\sm + \nm D \mu).
\end{align*}

\begin{algorithm}
\hrulefill
\begin{algorithmic}[1]
\For {$k \leftarrow 1$, $N_\text{iter}$}
   \State $\bar{\lambda}_1  = T^\top (u_\lambda + \sigma_\lambda (\ybl - y_\lambda))$ 
   \State $\bar{\lambda}_2  = \nl D^\top (v_\lambda + \sigma_\lambda (\zbl - z_\lambda))$ 
   \State $\lambda \leftarrow \mathcal{P}_{\text{simplex}(N_\text{total})}
           \left(\lambda - \tau_\lambda( \bar{\lambda}_1 + \bar{\lambda}_2) \right)$

   \State $\ybl = T\lambda$
   \State $\zbl = \nl D\lambda$
   \State $\bar{\mu}_1  = P^\top( u_\mu+ \sigma_\mu(\ybm - y_\mu))$
   \State $\bar{\mu}_2  = \nm D^\top( v_\mu+ \sigma_\mu(\zbm - z_\mu))$
   \State $\mu  \leftarrow \mu - \tau_\mu  (\bar{\mu}_1 +\bar{\mu}_2) $

   \State $\ybm = P\mu$ 
   \State $\zbm = \nm D\mu$ 

   \State $v_{\mu, \, \ell} = \sum_i C_{i \ell} - (u_{\mu, \, \ell} - \sm \ybml) \; \; \forall \ell$
   \For {$k^\prime \leftarrow 1$, $N_y$}
      \State $b_{i \ell} = \ulil + \sl \yblil - \exp(-\yml)
             \; \; \forall i,\ell$
      \State $\ylil = \left( b_{i \ell} + \sqrt{b^2_{i \ell} + 4 \sl C_{i \ell}} \right)/(2\sl)
             \; \; \forall i,\ell$ 

      \State $\yll = \sum_i \ylil \; \; \forall \ell$

      \State $ \ypml = 0 \;\; \forall \ell$
      \For {$k^{\prime \prime} \leftarrow 1$, $N_\text{newt}$} 
          \State $ \psi^{(1)}_\ell = -\exp (-\ypml) \cdot \yll +\sm \ypml + v_{\mu , \, \ell}$
          \State $ \psi^{(2)}_\ell = \exp (-\ypml) \cdot \yll +\sm \;\; \forall \ell$
          \State $ \ypml \leftarrow \ypml  - \psi^{(1)}_\ell \cdot \left( \psi^{(2)}_\ell\right)^{-1} \;\; \forall \ell$

         \State $ \ypml \leftarrow \pos (\ypml) \;\; \forall \ell$ 
      \EndFor
      \State $ \yml = \ypml \;\; \forall \ell$ 

   \EndFor

   \State $\zl \leftarrow \mathcal{P}_{L1(\nl \gamma_\lambda)}(\vl/\sl + \zbl)$
   \State $\zm \leftarrow \mathcal{P}_{L1(\nm \gamma_\mu)}(\vm/\sm + \zbm)$

   \State $\ul \leftarrow  \ul + \sl (\ybl  - \yl ) $
   \State $\um \leftarrow  \um + \sm (\ybm  - \ym ) $
   \State $\vl \leftarrow  \vl + \sl (\zbl  - \zl ) $
   \State $\vm \leftarrow  \vm + \sm (\zbm  - \zm ) $
\EndFor
\end{algorithmic}
\hrulefill
\caption{ADMM pseudocode for TV-constrained SAA estimation. Variables
$\lambda$, $\mu$, $y_\lambda$, $y_\mu$, $\zl$, $\zm$,$\ybl$, $\ybm$, $\zbl$, $\zbm$,
$\ul$, $\um$, $\vl$, and $\vm$ are initialized
to zero. Step size ratio parameters $\rho_\lambda$ and $\rho_\mu$ are determined in the
grid search described in Sec.~IIIA.
}
\end{algorithm}

\subsubsection*{ADMM pseudocode for TV-constrained SAA estimation}

The derivations in this appendix yield the additional steps that are needed to
formulate the pseudocode shown in Algorithm~2 from the ADMM-SAA pseudocode in
Algorithm~1.


\begin{thebibliography}{10}

\bibitem{kinahan1998attenuation}
P.~E. Kinahan, D.~W. Townsend, T.~Beyer, and D.~Sashin,
\newblock ``Attenuation correction for a combined 3{D} {PET/CT} scanner,''
\newblock {\em Med. Phys.}, vol. 25, pp. 2046--2053, 1998.

\bibitem{xia2011ultra}
T.~Xia, A.~M. Alessio, B.~De Man, R.~Manjeshwar, E.~Asma, and P.~E. Kinahan,
\newblock ``Ultra-low dose {CT} attenuation correction for {PET/CT},''
\newblock {\em Phys. Med. Biol.}, vol. 57, pp. 309--328, 2011.

\bibitem{burgos2014attenuation}
N.~Burgos, M.~J. Cardoso, K.~Thielemans, M.~Modat, S.~Pedemonte, J.~Dickson,
  A.~Barnes, R.~Ahmed, C.~J.Mahoney, J.~M. Schott, J.~S. Duncan, D.~Atkinson,
  S.~R. Arridge, B.~F. Hutton, and S.~Ourselin,
\newblock ``Attenuation correction synthesis for hybrid {PET}-{MR} scanners:
  application to brain studies,''
\newblock {\em IEEE Trans. Med. Imaging}, vol. 33, pp. 2332--2341, 2014.

\bibitem{osman2003respiratory}
M.~M. Osman, C.~Cohade, Y.~Nakamoto, and R.~L. Wahl,
\newblock ``Respiratory motion artifacts on {PET} emission images obtained
  using {CT} attenuation correction on {PET-CT},''
\newblock {\em Euro. J. Nuc. Med. Mol. Imaging}, vol. 30, pp. 603--606, 2003.

\bibitem{natterer1992attenuation}
F.~Natterer and H.~Herzog,
\newblock ``Attenuation correction in positron emission tomography,''
\newblock {\em Math. Method. Appl. Sci.}, vol. 15, pp. 321--330, 1992.

\bibitem{nuyts1999simultaneous}
J.~Nuyts, P.~Dupont, S.~Stroobants, R.~Benninck, L.~Mortelmans, and P.~Suetens,
\newblock ``Simultaneous maximum a posteriori reconstruction of attenuation and
  activity distributions from emission sinograms,''
\newblock {\em IEEE Trans. Med. Imaging}, vol. 18, pp. 393--403, 1999.

\bibitem{lewellen1998time}
T.~K. Lewellen,
\newblock ``Time-of-flight {PET},''
\newblock in {\em Semin. Nucl. Med.}, 1998, vol.~28, pp. 268--275.

\bibitem{defrise2012time}
M.~Defrise, A.~Rezaei, and J.~Nuyts,
\newblock ``Time-of-flight {PET} data determine the attenuation sinogram up to
  a constant,''
\newblock {\em Phys. Med. Biol.}, vol. 57, pp. 885--899, 2012.

\bibitem{defrise2008continuous}
M.~Defrise, V.~Panin, C.~Michel, and M.~E. Casey,
\newblock ``Continuous and discrete data rebinning in time-of-flight {PET},''
\newblock {\em IEEE Trans. Med. Imaging}, vol. 27, pp. 1310--1322, 2008.

\bibitem{panin2010restoration}
V.~Y. Panin, M.~Defrise, and M.~E. Casey,
\newblock ``Restoration of fine azimuthal sampling of measured {TOF} projection
  data,''
\newblock in {\em 2010 IEEE Nucl. Sci. Symp. Med. Imaging Conf.}, 2011, pp.
  3079--3084.

\bibitem{rezaei2012simultaneous}
A.~Rezaei, M.~Defrise, G.~Bal, C.~Michel, M.~Conti, C.~Watson, and J.~Nuyts,
\newblock ``Simultaneous reconstruction of activity and attenuation in
  time-of-flight {PET},''
\newblock {\em IEEE Trans Med. Imaging}, vol. 31, pp. 2224--2233, 2012.

\bibitem{cheng2020maximum}
L.~Cheng, T.~Ma, X.~Zhang, Q.~Peng, Y.~Liu, and J.~Qi,
\newblock ``Maximum likelihood activity and attenuation estimation using both
  emission and transmission data with application to utilization of {Lu}-176
  background radiation in {TOF PET},''
\newblock {\em Med. Phys.}, vol. 47, pp. 1067--1082, 2020.

\bibitem{salomon2010simultaneous}
A.~Salomon, A.~Goedicke, B.~Schweizer, T.~Aach, and V.~Schulz,
\newblock ``Simultaneous reconstruction of activity and attenuation for
  {PET/MRI},''
\newblock {\em IEEE Trans. Med. Imaging}, vol. 30, pp. 804--813, 2010.

\bibitem{wolf2013few}
P.~A. Wolf, J.~S. J{\o}rgensen, T.~G. Schmidt, and E.~Y. Sidky,
\newblock ``Few-view single photon emission computed tomography {(SPECT)}
  reconstruction based on a blurred piecewise constant object model,''
\newblock {\em Phys. Med. Biol.}, vol. 58, pp. 5629--5652, 2013.

\bibitem{zhang2016investigation}
Z.~Zhang, J.~Ye, B.~Chen, A.~E. Perkins, S.~Rose, E.~Y. Sidky, C.-M. Kao,
  D.~Xia, C.-H. Tung, and X.~Pan,
\newblock ``Investigation of optimization-based reconstruction with an
  image-total-variation constraint in {PET},''
\newblock {\em Phys. Med. Biol.}, vol. 61, pp. 6055--6084, 2016.

\bibitem{zhang2018optimization}
Z.~Zhang, S.~Rose, J.~Ye, A.~E. Perkins, B.~Chen, C.-M. Kao, E.~Y. Sidky, C.-H.
  Tung, and X.~Pan,
\newblock ``Optimization-based image reconstruction from low-count, list-mode
  {TOF-PET} data,''
\newblock {\em IEEE Trans. Biomed. Eng.}, vol. 65, pp. 936--946, 2018.

\bibitem{chambolle2011first}
A.~Chambolle and T.~Pock,
\newblock ``A first-order primal-dual algorithm for convex problems with
  applications to imaging,''
\newblock {\em J. Math. lmaging Vis.}, vol. 40, pp. 120--145, 2011.

\bibitem{sidky2012convex}
E.~Y. Sidky, J.~H. J{\o}rgensen, and X.~Pan,
\newblock ``Convex optimization problem prototyping for image reconstruction in
  computed tomography with the {C}hambolle--{P}ock algorithm,''
\newblock {\em Phys. Med. Biol.}, vol. 57, pp. 3065--3091, 2012.

\bibitem{barber2020convergence}
R.~F. Barber and E.~Y. Sidky,
\newblock ``Convergence for nonconvex {ADMM}, with applications to {CT}
  imaging,''
\newblock {\em arXiv preprint arXiv:2006.07278}, 2020.

\bibitem{boyd2011distributed}
S.~Boyd, N.~Parikh, E.~Chu, B.~Peleato, and J.~Eckstein,
\newblock ``Distributed optimization and statistical learning via the
  alternating direction method of multipliers,''
\newblock {\em Found. Trend. Mach. Learn.}, vol. 3, pp. 1--122, 2011.

\bibitem{schmidt2022addressing}
T.~G. Schmidt, B.~A. Sammut, R.~F. Barber, X.~Pan, and E.~Y. Sidky,
\newblock ``Addressing {CT} metal artifacts using photon-counting detectors and
  one-step spectral {CT} image reconstruction,''
\newblock {\em Med. Phys.}, vol. 49, pp. 3021--3040, 2022.

\bibitem{lange1990convergence}
K.~Lange,
\newblock ``Convergence of {EM} image reconstruction algorithms with {G}ibbs
  smoothing,''
\newblock {\em IEEE Trans. Med. Imaging}, vol. 9, pp. 439--446, 1990.

\bibitem{heusser2017mlaa}
T.~Heu{\ss}er, C.~M. Rank, Y.~Berker, M.~T. Freitag, and M.~Kachelrie{\ss},
\newblock ``{MLAA}-based attenuation correction of flexible hardware components
  in hybrid {PET/MR} imaging,''
\newblock {\em EJNMMI physics}, vol. 4, pp. 1--23, 2017.

\bibitem{mehranian2017mr}
A.~Mehranian, H.~Zaidi, and A.~J. Reader,
\newblock ``{MR}-guided joint reconstruction of activity and attenuation in
  brain {PET-MR},''
\newblock {\em Neuro{I}mage}, vol. 162, pp. 276--288, 2017.

\bibitem{chun2016joint}
S.~Y. Chun, K.~Y. Kim, J.~S. Lee, and J.~A. Fessler,
\newblock ``Joint estimation of activity distribution and attenuation map for
  {TOF-PET} using alternating direction method of multiplier,''
\newblock in {\em Proc. IEEE 13th Int. Symp. Biomed. Imaging (ISBI)}, 2016, pp.
  86--89.

\bibitem{nien2014fast}
H.~Nien and J.~A. Fessler,
\newblock ``Fast {X}-ray {CT} image reconstruction using a linearized augmented
  {L}agrangian method with ordered subsets,''
\newblock {\em IEEE Trans. Med. Imaging}, vol. 34, pp. 388--399, 2014.

\bibitem{barber2016}
R.~F. Barber, E.~Y. Sidky, T.~G. Schmidt, and X.~Pan,
\newblock ``An algorithm for constrained one-step inversion of spectral {CT}
  data,''
\newblock {\em Phys. Med. Biol.}, vol. 61, pp. 3784--3818, 2016.

\bibitem{duchi2008efficient}
J.~Duchi, S.~Shalev-Shwartz, Y.~Singer, and T.~Chandra,
\newblock ``Efficient projections onto the $\ell_1$-ball for learning in high
  dimensions,''
\newblock in {\em Proceedings of the 25th international conference on Machine
  learning}, 2008, pp. 272--279.

\bibitem{our_arXiv}
Z.~Ren, E.~Y. Sidky, R.~F. Barber, C.-M. Kao, and X.~Pan,
\newblock ``Simultaneous activity and attenuation estimation in {TOF-PET} with
  {TV}-constrained nonconvex optimization,''
\newblock {\em arXiv preprint arXiv:2303.17042}, 2024.

\bibitem{green1990bayesian}
P.~J. Green,
\newblock ``Bayesian reconstructions from emission tomography data using a
  modified {EM} algorithm,''
\newblock {\em IEEE Trans. Med. Imaging}, vol. 9, pp. 84--93, 1990.

\bibitem{pierce2015digital}
L.~A. Pierce, B.~F. Elston, D.~A. Clunie, D.~Nelson, and P.~E. Kinahan,
\newblock ``A digital reference object to analyze calculation accuracy of {PET}
  standardized uptake value,''
\newblock {\em Radiology}, vol. 277, pp. 538--545, 2015.

\end{thebibliography}

\end{document}